# Optimal Tests of Treatment Effects for the Overall Population and Two Subpopulations in Randomized Trials, using Sparse Linear Programming

Michael Rosenblum*, Han Liu†, and En-Hsu Yen‡

June 4, 2013

**Abstract**

We propose new, optimal methods for analyzing randomized trials, when it is suspected that treatment effects may differ in two predefined subpopulations. Such subpopulations could be defined by a biomarker or risk factor measured at baseline. The goal is to simultaneously learn which subpopulations benefit from an experimental treatment, while providing strong control of the familywise Type I error rate. We formalize this as a multiple testing problem and show it is computationally infeasible to solve using existing techniques. Our solution involves a novel approach, in which we first transform the original multiple testing problem into a large, sparse linear program. We then solve this problem using advanced optimization techniques. This general method can solve a variety of multiple testing problems and decision theory problems related to optimal trial design, for which no solution was previously available. In particular, we construct new multiple testing procedures that satisfy minimax and Bayes optimality criteria. For a given optimality criterion, our new approach yields the optimal tradeoff between power to detect an effect in the overall population versus power to detect effects in subpopulations. We demonstrate our approach in examples motivated by two randomized trials of new treatments for HIV.

*Department of Biostatistics, Johns Hopkins Bloomberg School of Public Health, Baltimore, MD, USA, 21205, mrosenbl@jhsph.edu
†Department of Operations Research and Financial Engineering, Princeton University, Princeton, NJ, USA, 08544
‡Intel-NTU Connected Context Computing Center, Taipei, Taiwan, 10617

# 1 Introduction

An important goal of health research is determining which populations, if any, benefit from new treatments. Randomized trials are generally considered the gold standard for producing evidence of treatment effects. Most randomized trials aim to determine how a treatment compares to control, on average, for a given population. This results in trials that may fail to detect important differences in benefits and harms for subpopulations, such as those with a certain biomarker or risk factor. This problem affects trials in virtually all disease areas.

Consider planning a randomized trial of an experimental treatment versus control, where there is prior evidence that treatment effects may differ for two, predefined subpopulations. Such evidence could be from past trials or observational studies, or from medical knowledge of how the treatment is conjectured to work. Our goal is to construct a multiple testing procedure with optimal power to detect treatment effects for the overall population and for each subpopulation. We consider both Bayes and minimax optimality criteria. Existing multiple testing procedures in general do not satisfy either of these criteria.

It is a challenging problem to construct optimal multiple testing procedures. According to Romano et al. (2011), "there are very few results on optimality in the multiple testing literature." The problems we consider are especially challenging since we require strong control of the familywise Type I error rate, also called the studywide Type I error rate, as defined by Hochberg and Tamhane (1987). That is, we require that under any data generating distribution, the probability of rejecting one or more true null hypotheses is at most a given level $\alpha$. We incorporate these constraints because control of the studywide Type I error rate is generally required by regulatory agencies such as the U.S. Food and Drug Administration and the European Medicines Agency for confirmatory randomized trials involving multiple hypotheses (FDA and EMEA, 1998).

Strong control of the familywise Type I error rate implies infinitely many constraints, i.e., one for every possible data generating distribution. The crux of our problem is constructing multiple testing procedures satisfying all these constraints and optimizing power at a given set of alternatives. In the simpler problem of testing only the null hypothesis for the overall population, the issue of infinitely many constraints can be sidestepped; this is because for most reasonable tests, strong control of the Type I error is implied by control of the Type I error at the global null hypothesis of zero average treatment effect. In contrast, when dealing with multiple populations, procedures that control the familywise Type I error at the global null hypothesis can have quite large Type I error at distributions corresponding to a positive effect for one subpopulation and a nonpositive effect for another. For this reason, optimization methods designed for a single null hypothesis, such as those of Jennison (1987);

Eales and Jennison (1992); Banerjee and Tsiatis (2006); and Hampson and Jennison (2013), do not directly apply to our problem. Though in principle these methods could be extended to handle more Type I error constraints, such extensions are computationally infeasible in our problems, as we discuss in Section 7.

Our solution hinges on a novel method for transforming a fine discretization of the original multiple testing problem into a large, sparse linear program. The resulting linear program typically has over a million variables and constraints. We tailor advanced optimization tools to solve the linear program. To the best of our knowledge, this is the first computationally feasible method for constructing Bayes or minimax optimal tests of treatment effects for subpopulations and the overall population, while maintaining strong control of the familywise Type I error rate.

We apply our approach to answer the following open questions: What is the maximum power that can be gained to detect treatment effects in subpopulations if one is willing to sacrifice $x\%$ power for detecting an effect in the overall population? What is the minimum additional sample size required to increase power for detecting treatment effects in subpopulations by $x\%$, while maintaining a desired power for the overall population?

A motivating data example is given in Section 2. We define our problem in Section 3, present our method for solving it in Section 4, and demonstrate this method in Section 5. We explain how we overcome computational challenges in our problem in Sections 6 and 7. Sections 8 and 9 give extensions of our method to decision theory and minimax problems. The sparse linear programming algorithm we use is given in Section 10. We conclude with a discussion of limitations of our approach and future directions for research in Section 11.

## 2 Example: Randomized Trials of New Antiretroviral Treatments for HIV

We demonstrate our approach in scenarios motivated by two recently completed randomized trials of maraviroc, an antiretroviral medication for treatment-experienced, HIV positive individuals (Fätkenheuer et al., 2008). There is suggestive evidence from these trials that the treatment benefit may differ depending on the suppressive effect of an individual's background therapy, as measured by the phenotypic sensitivity score (PSS) at baseline. The estimated average treatment benefit of maraviroc among individuals with PSS less than 3 was larger than that among individuals with PSS 3 or more. This pattern has been observed for other antiretroviral medications, e.g., in randomized trials of etravirine (Katlama et al., 2009). We refer to those with PSS less than 3 as subpopulation 1, and those with PSS 3 or more as subpopulation 2. In the combined maraviroc trials, 63% of participants are in subpopulation 1.

In planning a trial of a new antiretroviral medication, it may be of interest to determine the average treatment effect for the overall population and for each of these subpopulations. We construct multiple testing procedures that maximize power for detecting treatment benefits in each subpopulation, subject to constraints on the familywise Type I error rate and on power for the overall population.

## 3 Multiple Testing Problem

### 3.1 Null Hypotheses and Test Statistics

Consider a randomized trial comparing a new treatment (a=1) to control (a=0), in which there are two prespecified subpopulations that partition the overall population. Denote the fraction of the overall population in subpopulation $k \in \{1, 2\}$ by $p_k$. We assume each patient is randomized to the new treatment or control with probability 1/2, independent of the patient's subpopulation. Below, for clarity of presentation, we focus on normally distributed outcomes with known variances. In Section A of the Supplementary Materials, we describe asymptotic extensions allowing a variety of outcome types, and where the variances are unknown and must be estimated.

For each subpopulation $k \in \{1, 2\}$ and study arm $a \in \{0, 1\}$, assume the corresponding patient outcomes are independent and distributed as $Y_{ka,i} \sim N(\mu_{ka}, \sigma_{ka}^2)$, for each patient $i = 1, 2, \ldots, n_{ka}$. For each subpopulation $k \in \{1, 2\}$, define the population average treatment effect as $\Delta_k = \mu_{k1} - \mu_{k0}$. For each $k \in \{1, 2\}$, define $H_{0k}$ to be the null hypothesis $\Delta_k \leq 0$, i.e., that treatment is no more effective than control, on average, for subpopulation k; define $H_{0C}$ to be the null hypothesis $p_1 \Delta_1 + p_2 \Delta_2 \leq 0$, i.e., that treatment is no more effective than control, on average, for the combined population.

Let $n$ denote the total sample size in the trial. For each $k \in \{1, 2\}$ and $a \in \{0, 1\}$, we assume the corresponding sample size $n_{ka} = p_k n / 2$; that is, the proportion of the sample in each subpopulation equals the corresponding population proportion $p_k$, and exactly half of the participants in each subpopulation are assigned to each study arm. This latter property can be approximately achieved by block randomization within each subpopulation.

We assume the subpopulation fractions $p_k$ and the variances $\sigma_{ka}^2$ are known. This implies the following z-statistics are sufficient statistics for $(\Delta_1, \Delta_2)$:

$$\text{for each subpopulation } k \in \{1, 2\}, Z_k = \left( \frac{1}{n_{k1}} \sum_{i=1}^{n_{k1}} Y_{k1,i} - \frac{1}{n_{k0}} \sum_{i=1}^{n_{k0}} Y_{k0,i} \right) v_k^{-1/2},$$

for $v_k = \sigma_{k1}^2 / n_{k1} + \sigma_{k0}^2 / n_{k0}$.

We also consider the pooled z-statistic for the combined population,

$$Z_C = \sum_{k=1}^{2} p_k \left( \frac{1}{n_{k1}} \sum_{i=1}^{n_{k1}} Y_{k1,i} - \frac{1}{n_{k0}} \sum_{i=1}^{n_{k0}} Y_{k0,i} \right) \left( p_1^2 v_1 + p_2^2 v_2 \right)^{-1/2}.$$

We then have $Z_C = \rho_1 Z_1 + \rho_2 Z_2$, for $\rho_k = [p_k^2 v_k/(p_1^2 v_1 + p_2^2 v_2)]^{1/2}$, which is the covariance of $Z_k$ and $Z_C$. The vector of sufficient statistics $(Z_1, Z_2)$ is bivariate normal with mean

$$(\delta_1, \delta_2) = (\Delta_1/\sqrt{v_1}, \Delta_2/\sqrt{v_2}), \qquad (1)$$

and covariance matrix the identity matrix. We call $(\delta_1, \delta_2)$ the non-centrality parameters of $(Z_1, Z_2)$. For $\Delta^{\min} > 0$ the minimum, clinically meaningful treatment effect, let $\delta_1^{\min}$ and $\delta_2^{\min}$ be the non-centrality parameters that correspond to $\Delta_1 = \Delta^{\min}$ and $\Delta_2 = \Delta^{\min}$, respectively.

Define $\delta_C = EZ_C = \rho_1\delta_1 + \rho_2\delta_2$. We use the following equivalent representation of the null hypotheses above:

$$H_{01} : \delta_1 \leq 0; \qquad H_{02} : \delta_2 \leq 0; \qquad H_{0C} : \rho_1\delta_1 + \rho_2\delta_2 \leq 0. \qquad (2)$$

For any $(\delta_1, \delta_2)$, denote the corresponding set of true null hypotheses in the family $\mathcal{H} = \{H_{01}, H_{02}, H_{0C}\}$ by $\mathcal{H}_{\text{TRUE}}(\delta_1, \delta_2)$; for each $k \in \{1, 2\}$, this set contains $H_{0k}$ if and only if $\delta_k \leq 0$, and contains $H_{0C}$ if and only if $\rho_1\delta_1 + \rho_2\delta_2 \leq 0$.

### 3.2  Multiple Testing Procedures and Optimization Problem

The multiple testing problem is to determine which subset of $\mathcal{H}$ to reject, on observing a single realization of $(Z_1, Z_2)$. The pair $(Z_1, Z_2)$ is drawn from the distribution $P_{\delta_1,\delta_2}$, defined to be the bivariate normal distribution with mean vector $(\delta_1, \delta_2)$ and covariance matrix the $2 \times 2$ identity matrix.

Let $\mathcal{S}$ denote an ordered list of all subsets of the null hypotheses $\mathcal{H}$. Consider multiple testing procedures for the family of null hypotheses $\mathcal{H}$, i.e., maps from each possible realization of $(Z_1, Z_2)$ to an element of $\mathcal{S}$, representing the null hypotheses rejected upon observing $(Z_1, Z_2)$. It will be useful to consider the class $\mathcal{M}$ of randomized multiple testing procedures, defined as the maps $M$ from each possible realization of $(Z_1, Z_2)$ to a random variable taking values in $\mathcal{S}$. Formally, a randomized multiple testing procedure is a measurable map $M = M(Z_1, Z_2, U)$ that depends on $(Z_1, Z_2)$ but also may depend on an independent random variable $U$ that has a uniform distribution on $[0, 1]$. For conciseness, we often write "multiple testing procedure" instead of "randomized multiple testing procedure," with the understanding that we deal with the latter throughout.

Let $L$ denote a bounded loss function, where $L(s; \delta_1, \delta_2)$ represents the loss if the subset $s \subseteq \mathcal{H}$ is rejected when the true non-centrality parameters are $(\delta_1, \delta_2)$. In Section 5, we define a loss function that penalizes failure to reject each subpopulation null hypothesis when the corresponding average treatment effect is at least $\Delta^{\min}$. Our general method can be applied to any bounded loss function that can be numerically integrated with respect to $\delta_1, \delta_2$ by standard software with high precision. In particular, we allow $L$ to be non-convex in $(\delta_1, \delta_2)$, which is the case in all our examples.

We next state the Bayes version of our general optimization problem. Let $\Lambda$ denote a prior distribution on the set of possible pairs of non-centrality parameters $(\delta_1, \delta_2)$. We assume $\Lambda$ is a distribution with compact support on $(\mathbb{R}^2, \mathcal{B})$, for $\mathcal{B}$ a $\sigma$-algebra over $\mathbb{R}^2$.

**Constrained Bayes Optimization Problem:** For given $\alpha > 0, \beta > 0, \delta_1^{\min}, \delta_2^{\min}$, $L$, and $\Lambda$, find the multiple testing procedure $M \in \mathcal{M}$ minimizing

$$\int E_{\delta_1, \delta_2} \{L(M(Z_1, Z_2, U); \delta_1, \delta_2)\} d\Lambda(\delta_1, \delta_2), \qquad (3)$$

under the familywise Type I error constraints: for any $(\delta_1, \delta_2) \in \mathbb{R}^2$,

$$P_{\delta_1, \delta_2}[M \text{ rejects any null hypotheses in } \mathcal{H}_{\text{TRUE}}(\delta_1, \delta_2)] \leq \alpha, \qquad (4)$$

and the power constraint for the combined population:

$$P_{\delta_1^{\min}, \delta_2^{\min}}(M \text{ rejects } H_{0C}) \geq 1 - \beta. \qquad (5)$$

The objective function (3) encodes the expected loss incurred by the testing procedure $M$, averaged over the prior distribution $\Lambda$. The constraints (4) enforce strong control of the familywise Type I error rate.

The corresponding minimax optimization problem replaces the objective function (3) by

$$\sup_{(\delta_1, \delta_2) \in \mathcal{P}} E_{\delta_1, \delta_2} L(M(Z_1, Z_2, U); \delta_1, \delta_2), \qquad (6)$$

for $\mathcal{P}$ a subset of $\mathbb{R}^2$ representing the alternatives of interest.

# 4 Solution to Constrained Bayes Optimization Problem

The above constrained Bayes optimization problem is either very difficult or impossible to solve analytically, due to the continuum of Type I error constraints that must be satisfied. Our approach involves discretizing the constrained Bayes optimization problem. We approximate the infinite set of constraints (4) by a finite set of constraints, and restrict to multiple testing procedures that are constant over small rectangles. This transforms the constrained Bayes optimization problem, which is non-convex, into a large, sparse linear program that we solve using advanced optimization tools. In Section 6, we bound the approximation error in the discretization using the dual linear program; we apply this to show the approximation error is very small in all our examples.

We first restrict to the class of multiple testing procedures $\mathcal{M}_b \subset \mathcal{M}$ that reject no hypotheses outside the region $B = [-b, b] \times [-b, b]$ for a fixed integer $b > 0$. After determining the structure of approximately optimal procedures in $\mathcal{M}_b$, we build on this structure to generate approximately optimal procedures in the larger class $\mathcal{M}$, as described in Section B of the Supplementary Materials.

We next discretize the Type I error constraints (4), and restrict to a finite subset of them. Consider the set of pairs $G = \{(\delta_1, \delta_2) : \delta_1 = 0 \text{ or } \delta_2 = 0 \text{ or } \rho_1 \delta_1 + \rho_2 \delta_2 = 0\}$. These are the pairs of non-centrality parameters at which the first subpopulation has zero average benefit, the second subpopulation has zero average benefit, or the combined population has zero average benefit. For fixed $\tau = (\tau_1, \tau_2)$, we restrict to the familywise Type I error constraints (4) corresponding to the following discretization of $G$:

$$G_{\tau, b} = [\{(k\tau_1, 0) : k \in \mathbb{Z}\} \cup \{(0, k\tau_2) : k \in \mathbb{Z}\} \cup \{(\rho_2 k \tau_1, -\rho_1 k \tau_1) : k \in \mathbb{Z}\}] \bigcap B.$$

The next step is to define a subclass of multiple testing procedures that are constant over small rectangles. For each $k, k' \in \mathbb{Z}$, define the rectangle $R_{k,k'} = [k\tau_1, (k+1)\tau_1) \times [k'\tau_2, (k'+1)\tau_2)$. Let $\mathcal{R}$ denote the set of such rectangles in the bounded region $B$, i.e., $\mathcal{R} = \{R_{k,k'} : k, k' \in \mathbb{Z}, R_{k,k'} \subset B\}$. Define $\mathcal{M}_\mathcal{R}$ to be the subclass of multiple testing procedures $M \in \mathcal{M}_b$ that, for any $u \in [0, 1]$ and rectangle $r \in \mathcal{R}$, satisfy $M(z_1, z_2, u) = M(z'_1, z'_2, u)$ whenever $(z_1, z_2)$ and $(z'_1, z'_2)$ are both in $r$. For any procedure $M \in \mathcal{M}_\mathcal{R}$, its behavior is completely characterized by the finite set of values $\mathbf{m} = \{m_{rs}\}_{r \in \mathcal{R}, s \in \mathcal{S}}$, where

$$m_{rs} = P\{M(Z_1, Z_2, U) \text{ rejects the subset of null hypotheses } s \mid (Z_1, Z_2) \in r\}. \quad (7)$$

For any $r \in \mathcal{R}$, it follows that

$$\sum_{s \in \mathcal{S}} m_{rs} = 1, \text{ and } m_{rs} \geq 0 \text{ for any } s \in \mathcal{S}. \tag{8}$$

Also, for any set of real values $\{m_{rs}\}_{r \in \mathcal{R}, s \in \mathcal{S}}$ satisfying (8), there is a multiple testing procedure $M \in \mathcal{M}_\mathcal{R}$ satisfying (7), i.e., the procedure $M$ that rejects the subset of null hypotheses $s$ with probability $m_{rs}$ when $(Z_1, Z_2) \in r$.

The advantage of the above discretization is that if we restrict to procedures in $\mathcal{M}_\mathcal{R}$, the objective function (3) and constraints (4)-(5) in the constrained Bayes optimization problem are each linear functions of the variables $\mathbf{m}$. This holds even when the loss function $L$ is non-convex. To show (3) is linear in $\mathbf{m}$, first consider the term inside the integral in (3):

$$E_{\delta_1, \delta_2} L(M(Z_1, Z_2, U); \delta_1, \delta_2) \tag{9}$$
$$= \sum_{r \in \mathcal{R}, s \in \mathcal{S}} E_{\delta_1, \delta_2}[L(M(Z_1, Z_2, U); \delta_1, \delta_2) \mid M(Z_1, Z_2, U) = s, (Z_1, Z_2) \in r]$$
$$\times P_{\delta_1, \delta_2}[M(Z_1, Z_2, U) = s \mid (Z_1, Z_2) \in r] P_{\delta_1, \delta_2}[(Z_1, Z_2) \in r]$$
$$= \sum_{r \in \mathcal{R}, s \in \mathcal{S}} L(s; \delta_1, \delta_2) P_{\delta_1, \delta_2}[(Z_1, Z_2) \in r] m_{rs}. \tag{10}$$

The objective function (3) is the integral over $\Lambda$ of (9), which by the above argument equals

$$\sum_{r \in \mathcal{R}, s \in \mathcal{S}} \left\{ \int L(s; \delta_1, \delta_2) P_{\delta_1, \delta_2}[(Z_1, Z_2) \in r] d\Lambda(\delta_1, \delta_2) \right\} m_{rs}. \tag{11}$$

The constraints (4) and (5) can be similarly represented as linear functions of $\mathbf{m}$, as we show in Section C of the Supplementary Materials.

Define the discretized problem to be the constrained Bayes optimization problem restricted to procedures in $\mathcal{M}_\mathcal{R}$, and replacing the familywise Type I error constraints (4) by those corresponding to $(\delta_1, \delta_2) \in G_{\tau,b}$. The discretized problem can be expressed as:

**Sparse Linear Program Representing Discretization of Original Problem (3)-(5):**

For given $\alpha > 0, \beta > 0, \delta_1^{\min}, \delta_2^{\min}, \tau, b, L,$ and $\Lambda$, find the set of real values $\mathbf{m} = \{m_{rs}\}_{r \in \mathcal{R}, s \in \mathcal{S}}$ minimizing (11) under the constraints:

$$\text{for all } (\delta_1, \delta_2) \in G_{\tau,b}, \sum_{r \in \mathcal{R}} \sum_{s \in \mathcal{S}: s \cap \mathcal{H}_{\text{TRUE}}(\delta_1, \delta_2) \neq \emptyset} P_{\delta_1, \delta_2}[(Z_1, Z_2) \in r] m_{rs} \leq \alpha; \tag{12}$$

$$\sum_{r \in \mathcal{R}} \sum_{s \in \mathcal{S}: H_{0C} \in s} P_{\delta_1^{\min}, \delta_2^{\min}}[(Z_1, Z_2) \in r] m_{rs} \geq 1 - \beta; \tag{13}$$

$$\text{for all } r \in \mathcal{R}, \sum_{s \in \mathcal{S}} m_{rs} = 1; \tag{14}$$

$$\text{for all } r \in \mathcal{R}, s \in \mathcal{S}, m_{rs} \geq 0. \tag{15}$$

The constraints (12) represent the familywise Type I error constraints (4) restricted to $(\delta_1, \delta_2) \in G_{\tau,b}$ and $M \in \mathcal{M}_\mathcal{R}$; (13) represents the power constraint (5) restricted to $\mathcal{M}_\mathcal{R}$. We refer to the value of the Bayes objective function (11) evaluated at $\mathbf{m}$ as the Bayes risk of $\mathbf{m}$. Denote the optimal solution to the above problem as $\mathbf{m}^* = \{m_{rs}^*\}_{r \in \mathcal{R}, s \in \mathcal{S}}$, which through (7) characterizes the corresponding multiple testing procedure which we denote by $M^* \in \mathcal{M}_\mathcal{R}$.

The constraint matrix for the above linear program is quite sparse, that is, a large fraction of its elements are 0. This is because for any $r \in \mathcal{R}$ the constraint (14) has only $|\mathcal{S}|$ nonzero elements, and for any $r \in \mathcal{R}, s \in \mathcal{S}$, the constraint (15) has only 1 nonzero element. The power constraint (13) and the familywise Type I error rate constraints (12) generally have many nonzero elements, but there are relatively few of these constraints compared to (14) and (15). In the examples in the next section, there are about a hundred familywise Type I error constraints, while there are over a million constraints of the type (14) and (15).

The coefficients in (12) and (13) can be computed by evaluating the bivariate normal probabilities $P_{\delta_1, \delta_2}[(Z_1, Z_2) \in r]$. This can be done with high precision, essentially instantaneously, by standard statistical software such as the pmvnorm function in the R package mvtnorm. For each $r \in \mathcal{R}, s \in \mathcal{S}$, the term in curly braces in the objective function (11) can be computed by numerical integration over $(\delta_1, \delta_2) \in \mathbb{R}^2$ with respect to the prior distribution $\Lambda$. We give R code implementing this in the Supplementary Materials. The minimax version (6) of the optimization problem from Section 3.2 can be similarly represented as a large, sparse linear program, as we describe in Section 8. We show in Section 10 how to efficiently solve the resulting discretized problems using advanced optimization tools.

Allowing our multiple testing procedures to be randomized is crucial to our approach. Otherwise, the above linear program would be an integer program, where each $m_{rs}$ must be 0 or 1; integer programs are generally much more difficult to solve than linear programs.

## 5 Application to HIV Example in Section 2

### 5.1 Solution to Optimization Problem in Four Special Cases

We illustrate our method by solving special cases of the constrained Bayes optimization problem. We use a loss function $\tilde{L}$ that imposes a penalty of 1 unit for failing to reject the null hypothesis for each subpopulation when the average treatment effect is at least the minimum, clinically meaningful level in that subpopulation. Define $\tilde{L}(s; \delta_1, \delta_2) = \sum_{k=1}^{2} 1[\delta_k \geq \delta_k^{\min}, H_{0k} \notin s]$, where $1[C]$ is the indicator function taking value 1 if $C$ is true and 0 otherwise. In Section D of the Supplementary Materials, we consider other loss functions.

The risk corresponding to $\tilde{L}$ has an interpretation in terms of power to reject subpopulation null hypotheses. We define the power of a procedure to reject a null hypothesis $H \in \mathcal{H}$ as the probability it rejects at least $H$ (and possibly other null hypotheses). For any non-centrality parameters $\delta_1 \geq \delta_1^{\min}, \delta_2 < \delta_2^{\min}$ and any $M \in \mathcal{M}_\mathcal{R}$, the risk $E_{\delta_1,\delta_2}\tilde{L}(M(Z_1, Z_2, U); \delta_1, \delta_2)$ equals one minus the power of $M$ to reject $H_{01}$ under $(\delta_1, \delta_2)$; an analogous statement holds for subpopulation 2. For $\delta_1 \geq \delta_1^{\min}, \delta_2 \geq \delta_2^{\min}$, the risk equals the sum of one minus the power to reject each subpopulation null hypothesis.

We specify the following prior on the non-centrality parameters $(\delta_1, \delta_2)$: $\Lambda = \sum_{j=1}^{4} w_j \lambda_j$, where $\mathbf{w} = (w_1, w_2, w_3, w_4)$ is a vector of weights. Let $\lambda_1, \lambda_2, \lambda_3, \lambda_4$ be point masses at $(0,0), (\delta_1^{\min}, 0), (0, \delta_2^{\min})$, and $(\delta_1^{\min}, \delta_2^{\min})$, respectively. We consider two cases below. In the first, called the symmetric case, we set the subpopulation proportions $p_1 = p_2 = 1/2$ and use the symmetric prior $\Lambda_1$ defined by weights $\mathbf{w}^{(1)} = (0.25, 0.25, 0.25, 0.25)$. In the second, called the asymmetric case, we set $p_1 = 0.63$ and use the prior $\Lambda_2$ defined by weights $\mathbf{w}^{(2)} = (0.2, 0.35, 0.1, 0.35)$; this case is motivated by the example in Section 2, where subpopulation 1 is 63% of the total population and is believed to have a greater likelihood of benefiting from treatment than subpopulation 2. In Section D of the Supplementary Materials, we give examples using a continuous prior distribution on $\mathbb{R}^2$.

For each case, we solved the corresponding linear program using the algorithm in Section 10. The dimensions of the small rectangles in the discretization are set at $\tau = (0.02, 0.02)$, and we set $b = 5$. Each discretized linear program then has 1,506,006 variables and 1,757,113 constraints; all but 106 of the constraints are sparse. We give the precise structure of this linear program in Section 10.

We set $\alpha = 0.05$ and set each variance $\sigma_{ka}^2$ to be a common value $\sigma^2$. Let $M_{H_{0C}}^{\mathrm{UMP}}$ denote the uniformly most powerful test of the single null hypothesis $H_{0C}$ at level $\alpha$, which rejects $H_{0C}$ if $(Z_1, Z_2)$ is in the region $R_{\text{UMP}} = \{(z_1, z_2) : \rho_1 z_1 + \rho_2 z_2 > \Phi^{-1}(1-\alpha)\}$, for $\Phi$ the standard normal cumulative distribution function. To allow a direct comparison with $M_{H_{0C}}^{\text{UMP}}$, we set the total sample size $n$ equal to $n_{\min}$, defined to be the minimum sample size such that $M_{H_{0C}}^{\text{UMP}}$ has 90% power to reject $H_{0C}$ when the treatment benefit in both populations equals $\Delta^{\min}$. We round all results to two decimal places.

Consider the symmetric case. Let $\mathbf{m}_{\text{sym}}^*(1-\beta)$ denote the solution to the discretized problem at $H_{0C}$ power constraint $1-\beta$. For $1-\beta = 0.9$, any multiple testing procedure that satisfies the power constraint (5) and the familywise Type I error constraint (4) at the global null hypothesis $(\delta_1, \delta_2) = (0,0)$ must reject $H_{0C}$ whenever $(Z_1, Z_2) \in R_{\text{UMP}}$ and cannot reject any null hypothesis when $(Z_1, Z_2) \notin R_{\text{UMP}}$, except possibly on a set of Lebesgue measure zero; this follows from Theorem 3.2.1 of Lehmann and Romano (2005). Since this must hold for the optimal procedure $\mathbf{m}_{\text{sym}}^*(0.9)$, what remained to be determined is what regions in $R_{\text{UMP}}$ correspond to $\mathbf{m}_{\text{sym}}^*(0.9)$ rejecting $H_{01}, H_{02}$, both, or neither. The rejection regions for $\mathbf{m}_{\text{sym}}^*(0.9)$, computed using our method, are depicted in Figure 1a. For each subset of null hypotheses $s \in \mathcal{S}$, the region where $\mathbf{m}_{\text{sym}}^*(0.9)$ rejects $s$ is shown in a different color.

Consider weakening the $H_{0C}$ power constraint from $1-\beta = 0.9$ to $0.88$. The optimal solution $\mathbf{m}_{\text{sym}}^*(0.88)$ is shown in Figure 1b. Unlike $\mathbf{m}_{\text{sym}}^*(0.9)$, the procedure $\mathbf{m}_{\text{sym}}^*(0.88)$ has substantial regions outside $R_{\text{UMP}}$ where it rejects a single subpopulation null hypothesis. However, there is a small region in $R_{\text{UMP}}$ where $\mathbf{m}_{\text{sym}}^*(0.88)$ does not reject any null hypothesis. Also, in some parts of $R_{\text{UMP}}$ corresponding to one z-statistic being large and positive while the other is negative, $\mathbf{m}_{\text{sym}}^*(0.88)$ only rejects the null hypothesis corresponding to the large z-statistic, while $\mathbf{m}_{\text{sym}}^*(0.9)$ rejects both this and $H_{0C}$.

The optimal solutions $\mathbf{m}_{\text{sym}}^*(0.9)$ and $\mathbf{m}_{\text{sym}}^*(0.88)$ illustrate a tradeoff between power for $H_{0C}$ and for $H_{01}, H_{02}$, as shown in the first two columns of Table 1. For each procedure, the first row gives one minus the Bayes risk, which is a weighted sum of power under the three alternatives $(\delta_1^{\min}, 0), (0, \delta_2^{\min})$, and $(\delta_1^{\min}, \delta_2^{\min})$; these alternatives correspond to the treatment only benefiting subpopulation 1, only benefiting subpopulation 2, and benefiting both subpopulations, respectively, at the minimum, clinically meaningful level. The contributions from each of these are given in rows 2-4 of Table 1. There is no contribution from the alternative $(0,0)$ since the loss function $\tilde{L}$ is identically zero there.

The upshot is that using the procedure $\mathbf{m}_{\text{sym}}^*(0.88)$ in place of $\mathbf{m}_{\text{sym}}^*(0.9)$ involves sacrificing 2% power for $H_{0C}$ at $(\delta_1^{\min}, \delta_2^{\min})$, but gaining 11% power to reject $H_{01}$ at $(\delta_1^{\min}, 0)$ plus an identical increase in power to reject $H_{02}$ at $(0, \delta_2^{\min})$. We further discuss this tradeoff over a range of $\beta$ values in Section 5.2.

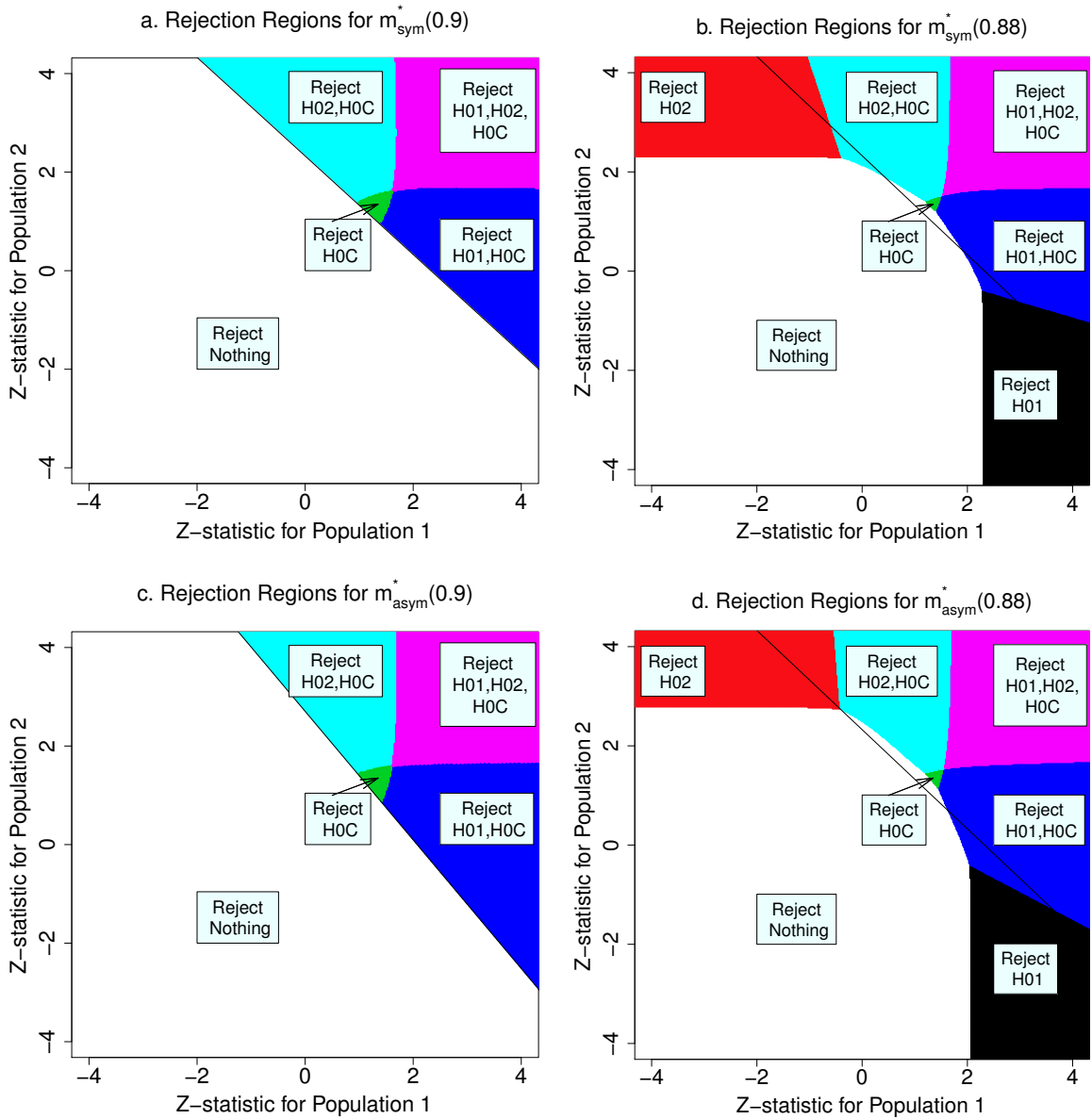

Figure 1: Optimal multiple testing procedures, for the symmetric case (a) and (b), and for the asymmetric case (c) and (d). In each plot, the black line is the boundary of $R_{\mathrm{UMP}}$.

Table 1: Bayes risk and power for optimal multiple testing procedures in symmetric and asymmetric cases, at $1-\beta = 0.9$ and $1-\beta = 0.88$.

|  | Symmetric Case | | Asymmetric Case | |
| --- | --- | --- | --- | --- |
|  | $\mathbf{m}^*_{\text{sym}}(0.9)$ | $\mathbf{m}^*_{\text{sym}}(0.88)$ | $\mathbf{m}^*_{\text{asym}}(0.9)$ | $\mathbf{m}^*_{\text{asym}}(0.88)$ |
| One Minus Bayes Risk | 0.52 | 0.58 | 0.67 | 0.71 |
| Power for $H_{01}$ at $(\delta_1^{\min}, 0)$ | 0.39 | 0.51 | 0.55 | 0.67 |
| Power for $H_{02}$ at $(0, \delta_2^{\min})$ | 0.39 | 0.51 | 0.25 | 0.30 |
| [Power $H_{01}$ at $(\delta_1^{\min}, \delta_2^{\min})$+ Power $H_{02}$ at $(\delta_1^{\min}, \delta_2^{\min})$]/2 | 0.65 | 0.66 | 0.64 | 0.64 |
| Power for $H_{0C}$ at $(\delta_1^{\min}, \delta_2^{\min})$ | 0.90 | 0.88 | 0.90 | 0.88 |

Next consider the asymmetric case, corresponding to $p_1 = 0.63$ and prior $\Lambda_2$. Let $\mathbf{m}^*_{\text{asym}}(1-\beta)$ denote the solution to the corresponding discretized problem at $1-\beta$. Figures 1c and 1d show the optimal solutions $\mathbf{m}^*_{\text{asym}}(0.9)$ and $\mathbf{m}^*_{\text{asym}}(0.88)$. The main difference between these and the solutions for the symmetric case is that $\mathbf{m}^*_{\text{asym}}(0.9)$ and $\mathbf{m}^*_{\text{asym}}(0.88)$ have larger rejection regions for $H_{01}$ and smaller rejection regions for $H_{02}$. The power tradeoff between $\mathbf{m}^*_{\text{asym}}(0.9)$ and $\mathbf{m}^*_{\text{asym}}(0.88)$ is given in the last two columns of Table 1. Sacrificing 2% power for $H_{0C}$ at $(\delta_1^{\min}, \delta_2^{\min})$ leads to an increase in 12% power to reject $H_{01}$ at $(\delta_1^{\min}, 0)$, and an increase in 5% power to reject $H_{02}$ at $(0, \delta_2^{\min})$.

Each of the multiple testing procedures above has desirable monotonicity properties. First consider $\mathbf{m}^*_{\text{sym}}(0.9)$. For any point $(z_1, z_2)$ for which $\mathbf{m}^*_{\text{sym}}(0.9)$ rejects $H_{01}$, it also rejects $H_{01}$ at any point $(z'_1, z_2) \in B$ for $z'_1 \geq z_1$. The analogous property holds for $H_{02}$. For any $(z_1, z_2)$ for which $\mathbf{m}^*_{\text{sym}}(0.9)$ rejects $H_{0C}$, it also rejects $H_{0C}$ at any point $(z'_1, z'_2) \in B$ for $z'_1 \geq z_1, z'_2 \geq z_2$. These monotonicity properties also hold for the other procedures above.

## 5.2 Optimal Power Tradeoff for Combined Population versus Subpopulations

We explore the tradeoffs in power for rejecting a subpopulation null hypothesis when the treatment only benefits one subpopulation, versus power for rejecting the combined population null hypothesis when the treatment benefits both subpopulations. Figure 2 shows the Bayes risk and its components for the optimal procedure $\mathbf{m}^*_{\text{sym}}(1-\beta)$, for each value of $1-\beta$ in a grid of points on the interval $[0.8, 0.9]$, for the symmetric case. The solid curve in Figure 2a gives the optimal tradeoff between the Bayes risk and the constraint $1-\beta$ on the power to reject $H_{0C}$ at $(\delta_1^{\min}, \delta_2^{\min})$. Figures 2b-d show the contribution to the Bayes risk from power under the three alternatives $(\delta_1^{\min}, 0), (0, \delta_2^{\min})$, and $(\delta_1^{\min}, \delta_2^{\min})$.

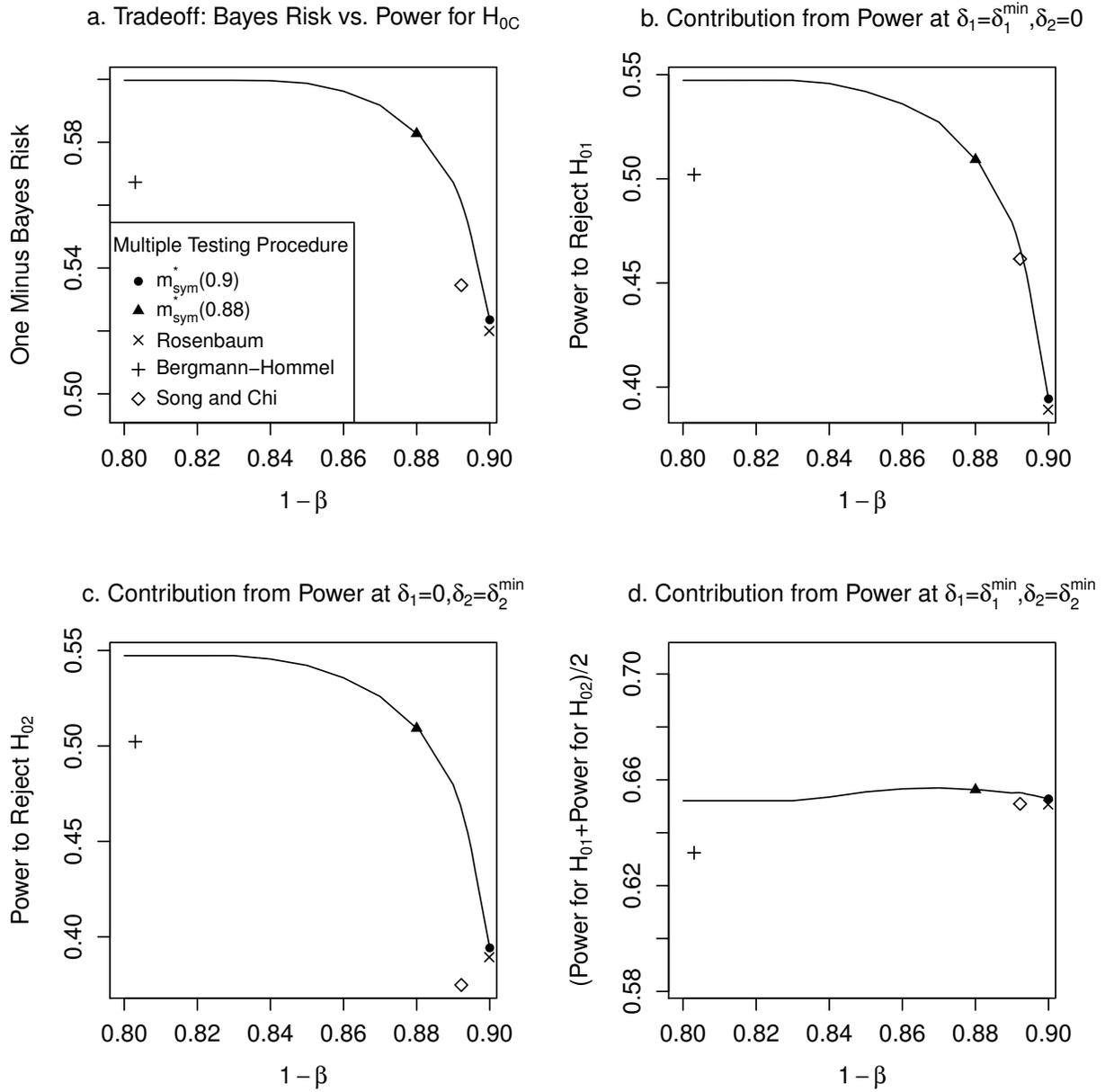

Figure 2: Optimal tradeoff between Bayes risk and power constraint $1 - \beta$ on $H_{0C}$, for symmetric case, i.e., $p_1 = p_2 = 1/2$ and prior $\Lambda_1$. In (a), we give one minus the Bayes risk on the vertical axis, so that in all four plots above, larger values represent better performance.

In each plot, we included points corresponding to $\mathbf{m}^*_{\text{sym}}(0.9)$ and $\mathbf{m}^*_{\text{sym}}(0.88)$, as well as three existing multiple testing procedures. The first is a procedure of Rosenbaum (2008) that rejects $H_{0C}$ when $M^{\text{UMP}}_{H_{0C}}$ does, and if so, additionally rejects each subpopulation null hypothesis $H_{0k}$ for which $Z_k > \Phi^{-1}(1-\alpha)$. The second existing method is an improvement on the Bonferroni and Holm procedures by Bergmann and Hommel (1988) for families of hypotheses that are logically related, as is the case here. The third is a special case of the method of Song and Chi (2007) that trades off power for $H_{0C}$ to increase power for $H_{01}$; we augmented their procedure to additionally reject $H_{02}$ in some cases. The details of the latter two procedures are given in Section E of the Supplementary Materials. Each of the three existing procedures strongly controls the familywise Type I error rate at level $\alpha$.

The procedure of Rosenbaum (2008) is quite close to the optimal threshold at $1-\beta = 0.9$, being suboptimal compared to $\mathbf{m}^*_{\text{sym}}(0.9)$ by only 0.4% in terms of the Bayes risk; the corresponding rejection regions are very similar to those of $\mathbf{m}^*_{\text{sym}}(0.9)$. The procedure of Bergmann and Hommel (1988) is suboptimal by 5% in power for rejecting $H_{01}$ at $(\delta_1^{\min}, 0)$ and for rejecting $H_{02}$ at $(0, \delta_2^{\min})$. The procedure of Song and Chi (2007) is close to optimal for rejecting $H_{01}$ at $(\delta_1^{\min}, 0)$, but is 9% suboptimal for $H_{02}$ at $(0, \delta_2^{\min})$. This is not surprising since their procedure was designed with a focus on the null hypothesis for a single subpopulation, rather than for both a subpopulation and its complement.

The tradeoff curves are steep near $1-\beta = 0.9$, indicating that a small sacrifice in power to reject $H_{0C}$ at $(\delta_1^{\min}, \delta_2^{\min})$ leads to a relatively large gain in power to detect subpopulation treatment effects when the treatment benefits only one subpopulation. The first two columns of Table 1, which compare $\mathbf{m}^*_{\text{sym}}(0.9)$ versus $\mathbf{m}^*_{\text{sym}}(0.88)$, are an example of this tradeoff. Diminishing returns set in for $1-\beta$ less than 0.84, in that there is negligible improvement in the Bayes risk or any of its components if one further relaxes the power constraint for $H_{0C}$.

Consider the impact of increasing the total sample size $n$ above $n_{\min}$, holding $\Delta^{\min}$ and the variances $\sigma^2_{sa}$ fixed. Define the multiple testing procedure $\mathbf{m}^*_{\text{SS}}(n)$ to be the solution to the discretized optimization problem in the symmetric case at $1-\beta = 0.9$ and sample size $n$, for $n \geq n_{\min}$. As $n$ increases from $n_{\min}$, the rejection regions of $\mathbf{m}^*_{\text{SS}}(n)$ progress from $\mathbf{m}^*_{\text{sym}}(0.9)$ as in Figure 1a to rejection regions qualitatively similar to $\mathbf{m}^*_{\text{sym}}(0.88)$ as in Figure 1b; these regions are given in Section F of the Supplementary Materials. Increasing sample size from $n = n_{\min}$ to $n = 1.06 n_{\min}$, the power of $\mathbf{m}^*_{\text{SS}}(n)$ to reject $H_{01}$ at $(\delta_1^{\min}, 0)$ increases from 42% to 52%; there is an identical increase in power to reject $H_{02}$ at $(0, \delta_2^{\min})$.

To give a sense of the value of increasing power from 42% to 52%, consider testing the single null hypothesis $H_{01}$ based on $Z_1$, using the uniformly most powerful test of $H_{01}$ at level $\alpha'$. Consider the sample size for which the power of this test is 42% at a fixed alternative $\Delta_1 = \Delta_1^{\min} > 0$ and $\sigma_{10}^2 > 0, \sigma_{11}^2 > 0$. To increase power to 52%, one needs to increase the sample size by 38%, 31%, or 28%, for $\alpha'$ equal to 0.05, 0.05/2 or 0.05/3, respectively. In light of this, the above 10% gains in power for detecting subpopulation treatment effects at the cost of only a 6% increase in sample size (and while maintaining 90% power for $H_{0C}$), as $\mathbf{m}_{SS}^*(n)$ does, is a relatively good bargain.

The tradeoff curve in Figure 2a is optimal, i.e., no multiple testing procedure satisfying the familywise Type I error constraints (4) can have Bayes risk and power for $H_{0C}$ corresponding to a point that exceeds this curve. The Bayes risk is a weighted combination of power at the three alternatives given above, as shown in Figures 2b-d. It follows that no multiple testing procedure satisfying (4) can simultaneously exceed all three power curves in Figures 2b-d. However, there do exist procedures that have power greater than one or two of these curves but that fall short on the other(s). By using different priors $\Lambda$, one can produce examples of such procedures. It is an area of future work to apply our method under a variety of priors to explore tradeoffs in power between $H_{01}$ and $H_{02}$.

A similar pattern as in Figure 2 holds for the asymmetric case. The main difference is that power to reject $H_{01}$ at $(\delta_1^{\min}, 0)$ is larger than power to reject $H_{02}$ at $(0, \delta_2^{\min})$.

In Section F of the Supplementary Materials, we answer the question posed in Section 1 of what minimum additional sample size is required to achieve a given power for detecting treatment effects in each subpopulation, while maintaining 90% power for $H_{0C}$ and strongly controlling the familywise Type I error rate. We do this for $p_1 = p_2$, but the general method can be applied to any subpopulation proportions.

# 6 Using the Dual of the Discretized Problem to Bound the Bayes Risk of the Original Problem

## 6.1 Active Constraints in the Dual Solution of the Discretized Problem

For each optimal procedure from Section 5.1, Figure 3 shows the constraints among (12) and (13) that are active, i.e., for which the corresponding inequalities hold with equality. In all cases, the global null hypothesis $(\delta_1, \delta_2) = (0, 0)$, the power constraint (13), and one constraint on the boundary of the null space for each of $H_{01}$ and $H_{02}$, are active. In addition, each of the optimal procedures at $1 - \beta = 0.88$ has two active constraints on the boundary of the null space for $H_{0C}$. The active familywise Type I error constraints correspond to the least-favorable distributions for a given procedure.

To illustrate the importance of all these constraints, consider what would happen if we only imposed the familywise Type I error constraint (4) at the global null hypothesis and the power constraint (5) at $1 - \beta = 0.88$. The optimal solution to the corresponding constrained Bayes optimization problem in the symmetric case has familywise Type I error 0.54 at non-centrality parameters $(\delta_1^{\min}, 0)$ and $(0, \delta_2^{\min})$; in the asymmetric case, the familywise Type I error at each of these alternatives is 0.39 and 0.69, respectively. The rejection regions are given in Section G of the Supplementary Materials. This demonstrates the importance of the additional familywise Type I error constraints.

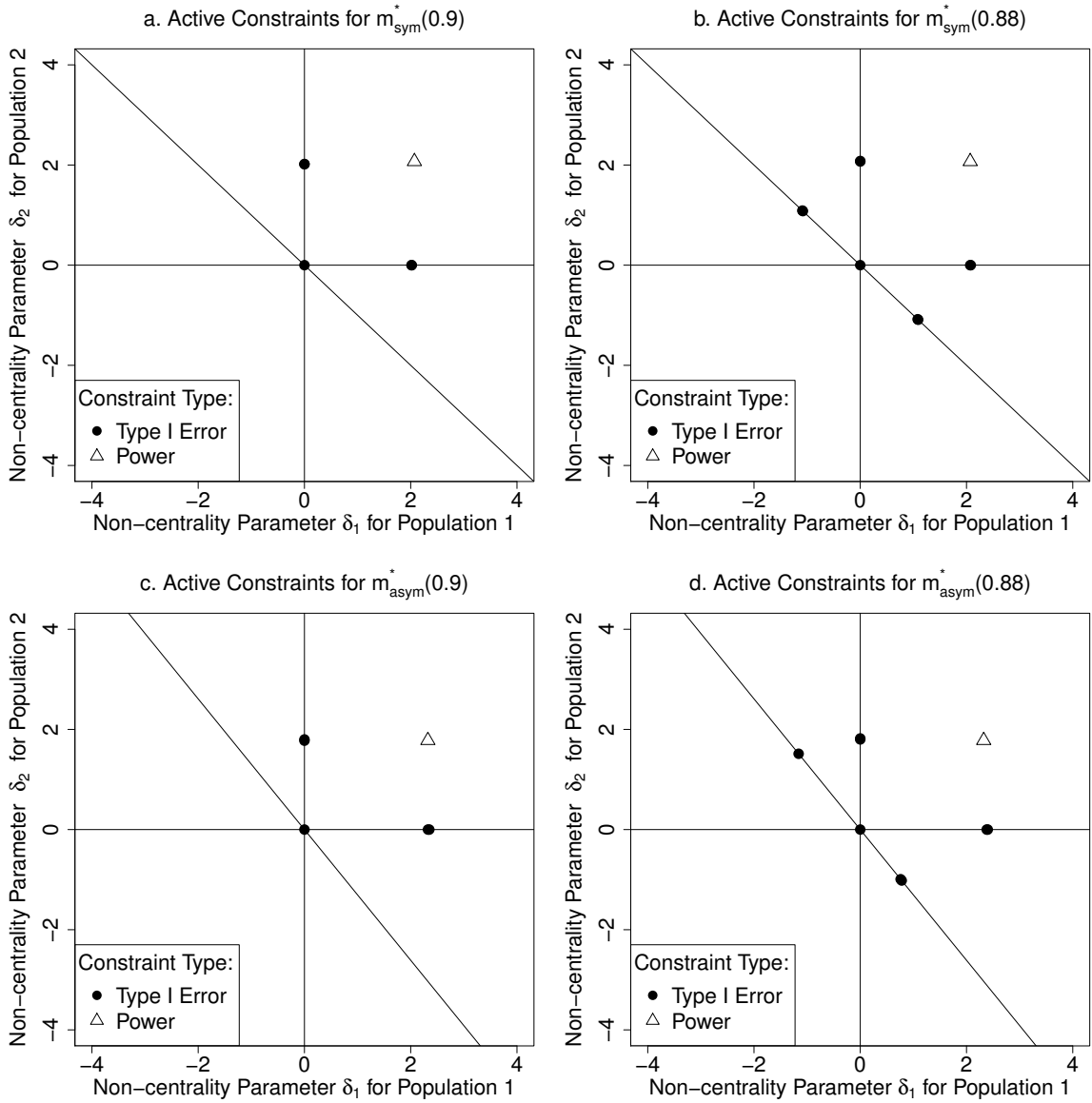

Figure 3: Active constraints for optimal procedures $\mathbf{m}^*_{\text{sym}}(0.9), \mathbf{m}^*_{\text{sym}}(0.88), \mathbf{m}^*_{\text{asym}}(0.9)$, and $\mathbf{m}^*_{\text{asym}}(0.88)$. Lines indicate boundaries of null spaces for $H_{01}, H_{02}, H_{0C}$.

## 6.2 Bounding the Bayes Risk of the Optimal Solution to the Original Problem

The optimal multiple testing procedures shown in Figure 1 are the solutions to versions of the discretized problem (11)-(15), which is an approximation to the constrained Bayes optimization problem (3)-(5). We refer to the latter as the original problem. Two natural questions are whether the solution to the discretized problem satisfies all constraints of the original problem, and how the optimal Bayes risk for the discretized problem compares to the optimal Bayes risk achievable in the original problem.

To answer the first question, we show in Section H of the Supplementary Materials that by solving the discretized problem setting $\alpha = 0.05 - 10^{-4}$ in the constraints (12), we ensure our procedures satisfy all constraints of the original problem at $\alpha = 0.05$. It remains to bound the difference between the optimal Bayes risk for the original problem and that for the discretized problem.

We use the optimal solution $\nu^*$ to the dual of the discretized problem to obtain a lower bound on optimal Bayes risk of the original problem. For a given discretized problem and optimal dual solution $\nu^*$, let $C_{\text{FWER}}$ denote the set of indices of active familywise Type I error constraints among (12); these are the indices $j$ of the pairs $(\delta_{1,j}, \delta_{2,j}) \in G_{\tau,b}$ corresponding to the nonzero components $\nu_j^*$ of $\nu^*$. Let $\nu_p^*$ denote the value of the dual variable corresponding to the power constraint (13). Let $\mathcal{M}_c$ denote the subclass of multiple testing procedures in $\mathcal{M}$ that satisfy all the constraints (4) and (5) of the original problem. Then we have the following lower bound on the objective function (3) of the original problem:

$$\inf_{M \in \mathcal{M}_c} \int E_{\delta_1,\delta_2} L(M(Z_1, Z_2, U); \delta_1, \delta_2) d\Lambda(\delta_1, \delta_2)$$

$$\geq \inf_{M \in \mathcal{M}} \left[ \int E_{\delta_1,\delta_2} L(M(Z_1, Z_2, U); \delta_1, \delta_2) d\Lambda(\delta_1, \delta_2) + \nu_p^* \left\{ 1 - \beta - P_{\delta_1^{\min}, \delta_2^{\min}}(M \text{ rejects } H_{0C}) \right\} \right.$$

$$\left. + \sum_{j \in C_{\text{FWER}}} \nu_j^* \left\{ P_{\delta_{1,j}, \delta_{2,j}} (M \text{ rejects any null hypothesis in } \mathcal{H}_{\text{TRUE}}(\delta_{1,j}, \delta_{2,j})) - \alpha \right\} \right], \quad (16)$$

which follows since all components of $\nu^*$ are nonnegative, by definition. The minimization problem (16) is straightforward to solve since it is unconstrained. We give the solution in Section I of the Supplementary Materials, which is computed by numerical integration. We then computed the absolute value of the difference between this lower bound and the Bayes risk of the optimal solution to the discretized problems in Section 5.1, which is at most 0.005 in each case. This shows the Bayes risk for the optimal solution to each discretized problem is within 0.005 of the optimum achievable in the original problem, so little is lost by restricting to the discretized procedures at the level of discretization we used.

# 7 Computational Challenge and Our Approach to Solving It

Previous methods, such as those of Jennison (1987); Eales and Jennison (1992); and Banerjee and Tsiatis (2006) are designed to test a null hypothesis for a single population. These methods require specifying one or two constraints that include the active constraints for a given problem. This can be done for a single population since often the global null hypothesis of zero treatment effect and a single power constraint suffice. However, as shown in the previous section, in our problem there can be 6 active constraints in cases of interest. Especially in the asymmetric case shown in Figure 3d, it would be very difficult to a priori guess this set of constraints or to do an exhaustive search over all subsets of 6 constraints in $G_{\tau,b}$. Even if the set of active constraints for a given problem were somehow known or correctly guessed, existing methods would still need to find values of dual variables $\nu^*$ for which the solution to the corresponding unconstrained problem (16) satisfies all the constraints with equality. This is computationally infeasible by exhaustive search or ad hoc optimization methods due to the dimension of the search.

Our approach overcomes the above computational obstacle by transforming a fine discretization of the original problem to a sparse linear program that contains many constraints; we then leverage the machinery of linear program solvers, which are expressly designed to optimize under many constraints simultaneously. The sparsity of the constraint matrix of the discretized linear program is crucial to the computational feasibility of our approach. This sparsity results from being able to a priori specify a subset $G_{\tau,b}$ of the familywise Type I error constraints that contains close approximations to the active constraints, where $G_{\tau,b}$ is not so large as to make the resulting linear program computationally intractable. The size of $G_{\tau,b}$ in the examples from Section 5.1 was 106, and in examples in the Supplementary Materials it was as large as 344. More generally our method is computationally feasible with $G_{\tau,b}$ having up to a thousand constraints.

In all our examples we narrowed the search for the active constraints by focusing on the boundaries of the null spaces for $H_{01}, H_{02}, H_{0C}$, and restricting to a fine grid of points on the union of these boundaries. We conjecture that this approach will work in many problems where there are three or four hypotheses of primary interest. We discuss issues with handling larger numbers of null hypotheses, which our method is not designed for and which are much more challenging in general, in Section 11.

## 8  Minimax Optimization Criterion

We can replace the Bayes objective function (3) by the minimax objective function (6), in which the maximum is taken over a finite set of alternatives $\mathcal{P}$. The resulting optimization problem can be solved by binary search over candidate values $v$ for (6), where at each step we compute whether there exists a solution to the set of constraints (4)-(5) plus the additional constraints $E_{\delta_1, \delta_2} L(M(Z_1, Z_2, U); \delta_1, \delta_2) \leq v$ for each $(\delta_1, \delta_2) \in \mathcal{P}$. For values $v$ where a solution exists, (6) must be less than or equal to $v$; conversely, if no solution exists, (6) must be greater than $v$. Each step of the binary search is done using the discretized constraints (12)-(15) from Section 4 plus the additional constraint

$$\sum_{r \in \mathcal{R}, s \in \mathcal{S}} L(s; \delta_1, \delta_2) P_{\delta_1, \delta_2}[(Z_1, Z_2) \in r] m_{rs} \leq v,$$

for each $(\delta_1, \delta_2) \in \mathcal{P}$. Determining whether a solution exists to a set of sparse linear constraints can be solved using a similar algorithm as in Section 10. In Section J of the Supplementary Materials, we apply this to minimax versions of the problems from Section 5.1.

## 9  Application to Decision Theory Framework

A drawback of the hypothesis testing framework when considering subpopulations is that it does not directly translate into clear treatment recommendations. For example, if the null hypotheses $H_{0C}$ and $H_{01}$ are rejected, it is not clear whether to recommend the treatment to subpopulation 2. We propose a decision theory framework that formalizes the goal of recommending treatments to precisely the subpopulations who benefit at a clinically meaningful level. Though in practice treatment recommendations take many factors into account, the proposed framework allows one to explore the tradeoffs in prioritizing different types of errors in treatment recommendations to different subpopulations. The resulting optimization problems, which were not solvable previously, are solved using our general approach.

We use the definitions in Section 3.1. Our goal is to construct a decision procedure $D$, i.e., a measurable map from any possible realization of $(Z_1, Z_2)$ to a set of subpopulations $(\emptyset, \{1\}, \{2\}, \text{ or } \{1, 2\})$ to recommend the new treatment to. We consider randomized de-cision procedures, i.e., we allow $D$ to additionally depend on a random variable $U$ that is independent of $Z_1, Z_2$ and that has uniform distribution on $[0, 1]$.

We next define a class of loss functions. For each subpopulation $k \in \{1, 2\}$, let $l_{k,FP}$ be a user-defined penalty for recommending the treatment to subpopulation $k$ when $\delta_k < \delta_k^{\min}$ (a False Positive); let $l_{k,FN}$ be the penalty for failing to recommend the treatment to subpopulation $k$ when $\delta_k \geq \delta_k^{\min}$ (a False Negative). Define the loss function $L_D(d; \delta_1, \delta_2) = L_{D,1}(d; \delta_1, \delta_2) + L_{D,2}(d; \delta_1, \delta_2)$, where for each $d \subseteq \{1, 2\}$ and $k \in \{1, 2\}$, $L_{D,k}(d; \delta_1, \delta_2) = l_{k,FP} 1[\delta_k < \delta_k^{\min}, k \in d] + l_{k,FN} 1[\delta_k \geq \delta_k^{\min}, k \notin d]$. For illustration, we consider two loss functions. The first, $L_D^{(1)}$, is defined by $l_{k,FN} = 1$ and $l_{k,FP} = 2$ for each $k$; the second, $L_D^{(2)}$, is defined by $l_{k,FN} = 2$ and $l_{k,FP} = 1$ for each $k$.

We minimize the following Bayes criterion analogous to (3):

$$\int E_{\delta_1, \delta_2} \{L(D(Z_1, Z_2, U); \delta_1, \delta_2)\} d\Lambda(\delta_1, \delta_2), \qquad (17)$$

over all decision procedures $D$ as defined above, under the following constraints:

$$\text{for any } (\delta_1, \delta_2) \in \mathbb{R}^2, \quad P_{\delta_1, \delta_2}\left[\sum_{k \in D(Z_1, Z_2, U)} p_k \Delta_k \leq 0\right] \leq \alpha. \qquad (18)$$

These constraints impose a bound of $\alpha$ on the probability of recommending the new treatment to an aggregate population (defined as the corresponding single subpopulation if $D = \{1\}$ or $\{2\}$, or the combined population if $D = \{1, 2\}$) having no average treatment benefit.

We consider the symmetric case from Section 5.1. The optimal decision regions are given in Figure 4. The optimal decision rule under $L_D^{(1)}$, denoted by $D^{(1)*}$, is more conservative in recommending the treatment than the optimal rule under $L_D^{(2)}$, denoted by $D^{(2)*}$. This is because the former loss function penalizes more for false positive recommendations. Table 2 contrasts $D^{(1)*}$ and $D^{(2)*}$. When $(\delta_1, \delta_2) = (\delta_1^{\min}, \delta_2^{\min})$, the conservative rule $D^{(1)*}$ recommends treatment to both subpopulations 21% less often compared to $D^{(2)*}$. However, when the treatment only benefits one subpopulation, the conservative rule $D^{(1)*}$ has 11% greater accuracy in recommending it to just that subpopulation.

One may prefer to strengthen (18) to require $P_{\delta_1, \delta_2}[D(Z_1, Z_2, U) \cap \{k : \delta_k \leq 0\} \neq \emptyset] \leq \alpha$, that is, to require probability at most $\alpha$ of recommending the new treatment to any subpopulation having no average treatment benefit. Our framework allows computation of the tradeoff between optimal procedures under these different sets of constraints, which is an area of future research.

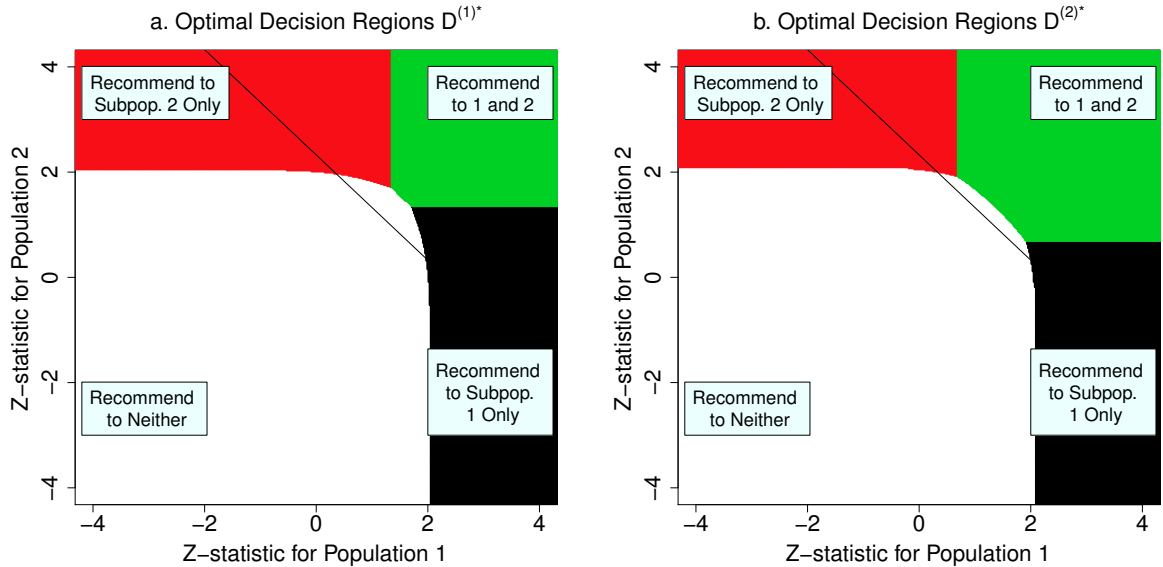

Figure 4: Optimal decision regions for the symmetric case ($p_1 = 1/2$ and prior $\Lambda_1$), using loss function (a) $L_D^{(1)}$ and (b) $L_D^{(2)}$. To allow comparability to Figure 1, we included the solid line, which is the boundary of the rejection region for $M_{H_{0C}}^{\text{UMP}}$.

Table 2: Probabilities of Different Recommendations by Optimal Decision Procedures $D^{(1)*}$ and $D^{(2)*}$, at three alternatives (Alt). The optimal recommendation (Rec.) at each alternative is in bold type.

| Alt: | $(\delta_1, \delta_2) = (\delta_1^{\min}, 0)$ | | | | $(\delta_1, \delta_2) = (\delta_1^{\min}, \delta_2^{\min})$ | | | | $(\delta_1, \delta_2) = (0, 0)$ | | | |
|---|---|---|---|---|---|---|---|---|---|---|---|---|
| Rec.: | $\emptyset$ | **{1}** | {2} | {1,2} | $\emptyset$ | {1} | {2} | **{1,2}** | **$\emptyset$** | {1} | {2} | {1,2} |
| $D^{(1)*}$ | 0.45 | **0.48** | 0.01 | 0.06 | 0.16 | 0.14 | 0.14 | **0.57** | **0.95** | 0.02 | 0.02 | 0.01 |
| $D^{(2)*}$ | 0.46 | **0.37** | 0 | 0.17 | 0.13 | 0.4 | 0.4 | **0.78** | **0.95** | 0.01 | 0.01 | 0.02 |

## 10 Algorithm to Solve Our Large, Sparse Linear Programs

The discretized problem from Section 4 can be represented as a large-scale linear programming problem. To show this, define the following ordering of subsets of $\mathcal{H}$:

$$\mathcal{S}' = (s_0, \ldots, s_6) = (\emptyset, \{H_{01}\}, \{H_{02}\}, \{H_{0C}\}, \{H_{01}, H_{0C}\}, \{H_{02}, H_{0C}\}, \{H_{01}, H_{02}, H_{0C}\}).$$

We leave out the subset $\{H_{01}, H_{02}\}$, since by the results of Sonnemann and Finner (1988) it suffices to consider only coherent multiple testing procedures, which in our context are those that reject $H_{0C}$ whenever $\{H_{01}, H_{02}\}$ is rejected. For a given ordering $r_1, r_2, \ldots$ of the rectangles $\mathcal{R}$, define $\mathbf{x} = (m_{r_1 s_1}, \ldots, m_{r_1 s_6}, m_{r_2 s_1}, \ldots, m_{r_2 s_6}, m_{r_3 s_1} \ldots)$, which has $n_v = |\mathcal{R}|(|\mathcal{S}'| - 1)$ components. We do not include the variables $m_{r_i s_0}$ in $\mathbf{x}$, since by (14) these variables are functions of variables already in $\mathbf{x}$; in particular, $m_{r_i s_0} = 1 - \sum_{j=1}^{6} m_{r_i s_j}$.

The discretized problem from Section 4 can be expressed in the canonical form:

$$\max_{x \in \mathbb{R}^{n_v}} \mathbf{c}^\mathbf{T} \mathbf{x} \quad \text{s.t.} \quad \mathbf{A} \mathbf{x} \leq \mathbf{b}. \tag{19}$$

The objective function $\mathbf{c}^\mathbf{T} \mathbf{x}$ represents the Bayes objective function (11). We set the first $n_d = |G_{\tau,b}| + 1$ rows of $\mathbf{A}$ to comprise the dense constraints, which include the familywise Type I error constraints (12) and the $H_{0C}$ power constraint (13). The remaining $n_s$ rows of $\mathbf{A}$ comprise the sparse constraints (14) and (15). Since $|\mathcal{R}| = (2b/\tau + 1)^2$, for the symmetric case in Section 5.1 with $b = 5$, $\tau = 0.02$, and $|\mathcal{S}'| - 1 = 6$, we have $n_v = |\mathcal{R}|(|\mathcal{S}'| - 1) = 1{,}506{,}006$, $n_d = |G_{\tau,b}| + 1 = 106$, and $n_s = |\mathcal{R}| + n_v = 1{,}757{,}007$. We then have

$\mathbf{A}$ is a $1{,}757{,}113 \times 1{,}506{,}006$ matrix with the following structure:

$$\mathbf{A} = \begin{bmatrix} n_d = 106 \text{ rows in which most elements are non-zero} \\ |\mathcal{R}| = 251{,}001 \text{ rows of the form:} \left\{ \begin{array}{l} 111111000000000000000000000000\ldots \\ 000000111111000000000000000000\ldots \\ 000000000000111111000000000000\ldots \\ \vdots \end{array} \right. \\ -\mathbf{I}_{1{,}506{,}006} \text{ , that is, the negative } n_v \times n_v \text{ identity matrix} \end{bmatrix},$$

$\mathbf{b}$ is a vector with $n_d + n_s = 1{,}757{,}113$ components (comp.) as follows:

$$\mathbf{b}^T = \begin{pmatrix} |G_{\tau,b}| = 105 \text{ comp.} & 1 \text{ comp.} & |\mathcal{R}| = 251{,}001 \text{ comp.} & n_v = 1{,}506{,}006 \text{ comp.} \\ \alpha, \alpha, \ldots, \alpha, & -(1-\beta), & 1, 1, \ldots, 1 & 0, 0, \ldots, 0 \end{pmatrix},$$

and $\mathbf{c}$ is a vector with $n_v = 1,506,006$ components.

The problem scale of (19) is quite large. In particular, the constraint matrix $\mathbf{A}$ has $\approx 2.6 \times 10^{12}$ entries. However, we can solve (19) by exploiting the sparsity structure of $\mathbf{A}$. We use a projected subgradient descent method, which consists of a subgradient descent step and a projection step, where the solution at iteration $k+1$ is

$$\mathbf{x}^{(k+1)} = P_s\left(\mathbf{x}^{(k)} - \delta_k \mathbf{g}^{(k)}\right), \tag{20}$$

where $P_s(.)$ means projection onto the feasible region determined by the sparse constraints, $\delta_k$ is a step size, and $\mathbf{g}^{(k)}$ is the subgradient of $\mathbf{x}_k$, defined as

$$\mathbf{g}^{(k)} = \begin{cases} \mathbf{c}, & \text{if for all } i = 1, \ldots, n_d,\ \mathbf{a}_i^T \mathbf{x}^{(k)} \leq b_i, \\ -\mathbf{a}_{i'}, & \text{otherwise, where } i' \text{ is a randomly selected index in } \{i : \mathbf{a}_i^T \mathbf{x}^{(k)} > b_i\}. \end{cases} \tag{21}$$

The projection operator $P_s(.)$ can be applied in $\mathcal{O}(n_v)$ floating point operations (flops) by computing the projection in $|\mathcal{R}|$ independent subsystems, each with $|\mathcal{S}'|-1$ variables. Checking violations of $n_d$ dense constraints together with the projection costs at most $\mathcal{O}(n_v(n_d+1))$ flops per iteration. The procedure (20) is guaranteed to converge to the optimum of (19) (Boyd et al., 2004). However, it may take a large number of iterations to achieve a high precision solution. In our implementation, we continue until an iteration $k'$ is reached where the proportion improvement in the objective function value is smaller than $10^{-3}$; we then use $\mathbf{x}^{(k')}$ as the initial point in a parametric simplex solver (Vanderbei, 2010). Though each iteration of a parametric simplex solver runs in superlinear time, for our problem it only requires a few iterations to move from $\mathbf{x}^{(k')}$ to a very precise optimal solution. Our solutions all had duality gap at most $10^{-8}$ showing they are within $10^{-8}$ of the true optimal solution to the discretized problem.

## 11 Discussion

Our method can be used with any bounded loss function that can be numerically integrated by standard software with high precision. This affords flexibility in specifying what penalties to impose for rejecting each possible subset of null hypotheses under each alternative $(\delta_1, \delta_2)$. As an example, we consider a loss function where the penalty for failing to reject each subpopulation null hypothesis is proportional to the magnitude of the treatment benefit in that subpopulation, in Section D of the Supplementary Materials.

An area of future research is to apply our methods to construct optimal testing proce-dures for trials comparing more than two treatments. Other potential applications include optimizing seamless Phase II/Phase III designs and adaptive enrichment designs.

We propose optimal methods for analyzing randomized trials when it is suspected that treatment effects may differ in two predefined subpopulations. It may be possible to extend our approach to three or four subpopulations, and this is an area for future research. However, with more than this many populations, our approach will likely be computationally infeasible. This is because a rate limiting factor is the number of variables in the discretized linear program. This number grows with the fineness of the discretization as well as the number of components in the sufficient statistic for the problem; our problem has only two components $(Z_1, Z_2)$, but for larger numbers of subpopulations this number would increase. It is a much more challenging (though still very important) problem to optimize tests and decision procedures for larger numbers of subpopulations.

Though the discretized problem involved optimizing over the class of randomized multiple testing procedures $\mathcal{M}_\mathcal{R}$, the optimal solutions in all our examples were deterministic procedures, i.e., each $m_{rs}^*$ was either 0 or 1. This is interesting, since there is no a priori guarantee that there exists an optimal deterministic solution, since the problem involves the large class of constraints (12).

In Section A of the Supplementary Materials, we show how our method can be extended to handle two-sided hypothesis tests.

## Acknowledgements

This research and analysis was supported by contract number HHSF2232010000072C, entitled, "Partnership in Applied Comparative Effectiveness Science," sponsored by the Food and Drug Administration, Department of Health and Human Services. This publication's contents are solely the responsibility of the authors and do not necessarily represent the official views of the above agencies.

**Supplementary Materials for "Optimal Tests of Treatment Effects for the Overall Population and Two Subpopulations in Randomized Trials, using Sparse Linear Programming"**

**Table of Contents**                                         **Pages**

## A  Extensions to Other Outcomes, Estimated Variances, and Two-sided Tests

We first generalize to outcomes other than normally distributed outcomes. Similar to Section 3.1, we assume that for each subject $i$, conditioned on his/her subpopulation $k \in \{1, 2\}$ and study arm assignment $a \in \{0, 1\}$, his/her data $Y_{ka,i}$ is a random draw from an unknown distribution $Q_{ka}$, and that this draw is independent of the data of all the other subjects. Instead of restricting each $Q_{ka}$ to be a normal distribution as in Section 3.1, we allow $Q_{ka}$ to be any distribution on $\mathbb{R}$. This allows, for example, the outcome to be binary valued, count valued, or continuous valued. Define $\mu_{ka}$ and $\sigma_{ka}^2$ to be the mean and variance, respectively, of $Q_{ka}$, which we assume to be finite. Let $\Delta_k = \mu_{k1} - \mu_{k0}$ for each $k \in \{1, 2\}$. The null hypotheses are as in Section 3.1 except using the above definition of $\Delta_k$. The z-statistics are as in Section 3.1, except using the outcomes $Y_{ka,i}$ which are distributed as $Q_{ka}$. The setup in Section 3.1 is then a special case of the above setup, if we let each $Q_{ka}$ be a normal distribution with mean $\mu_{ka}$ and variance $\sigma_{ka}^2$.

We assume each $\sigma_{ka}^2 > 0$. Then by the multivariate central limit theorem, the joint distribution of $(Z_1 - EZ_1, Z_2 - EZ_2, Z_C - EZ_C)$ converges to a zero mean, multivariate normal distribution with covariance matrix

$$\Sigma = \begin{pmatrix} 1 & 0 & \rho_1 \\ 0 & 1 & \rho_2 \\ \rho_1 & \rho_2 & 1 \end{pmatrix},$$

as sample size $n$ goes to infinity (holding $p_1, p_2$ and each $Q_{ka}$ fixed). The covariance matrix $\Sigma$ is the same as that for the case of normally distributed outcomes in Section 3.1. Therefore, the optimal multiple testing procedures constructed above for normally distributed outcomes can be expected to perform similarly for the more general case above, at large enough sample sizes. By a similar argument, when replacing variances by sample variances in the definition of the z-statistics, one expects the same to hold at large sample sizes. The sample size at which the normal approximation is a good one depends on features of the data generating distributions $Q_{ka}$. It is an area of future work to explore the impact of, e.g., skewed and heavy-tailed distributions $Q_{ka}$ on the performance of our multiple testing procedures.

A more formal argument would require consideration of local alternatives, that is, sequences of data generating distributions $Q_{ka}^{(n)}$ with $\Delta_k$ of order $1/\sqrt{n}$; this is because at fixed alternatives $Q_{ka}$ with $\Delta_k \neq 0$, the absolute value of the non-centrality parameter $\delta_k$ converges to infinity and so all reasonable procedures have power converging to 1. Extending our results under local alternatives is an area of future work.

We now consider null hypotheses related to two-sided tests. For each $k \in \{1, 2\}$, define $H'_{0k}$ to be the null hypothesis $\Delta_k = 0$, i.e., that treatment is equally as effective as control, on average, for subpopulation $k$; define $H'_{0C}$ to be the null hypothesis $p_1\Delta_1 + p_2\Delta_2 = 0$, i.e., that treatment is equally as effective as control, on average, for the combined population. Our general method can be applied to these hypotheses. One can use the same set of discretized constraints $G_{\tau,b}$, since the boundaries of the null hypotheses $\mathcal{H}$ and the boundaries of the null hypotheses $\{H'_{01}, H'_{02}, H'_{0C}\}$ are identical. The main changes would be that one may decide to specify a power constraint as in (5) not only at $(\delta_1^{\min}, \delta_2^{\min})$, but also at $(-\delta_1^{\min}, -\delta_2^{\min})$; one may also decide to assign weight in the prior to alternatives $(\delta_1, \delta_2)$ with $\delta_1$ and/or $\delta_2$ negative.

## B  Extending the Rejection Regions of a Solution to the Discretized Problem

Recall we restricted to the class of multiple testing procedures $\mathcal{M}_b \subset \mathcal{M}$ that reject no hypotheses outside the region $B = [-b, b] \times [-b, b]$ for a fixed integer $b > 0$. The reason was to make our solution computationally feasible. We now show how to iteratively augment the optimal solution to the discretized problem within the class $\mathcal{M}_\mathcal{R}$, which we denote by $M_b^*$, to allow rejection of null hypotheses outside $B = [-b, b] \times [-b, b]$. Let $B' = [-b', b'] \times [-b', b']$ for an integer $b' > b$. Define $\mathcal{R}' = \{R_{k,k'} : k, k' \in \mathbb{Z}, R_{k,k'} \subset B' \setminus B\}$. Let $\mathcal{M}'$ denote the class of multiple testing procedures $M \in \mathcal{M}_{b'}$ such that for any $u \in [0, 1]$, we have (i) $M(z_1, z_2, u) = M_b^*(z_1, z_2, u)$ for any $(z_1, z_2) \in B$, and (ii) for any rectangle $r \in \mathcal{R}'$ we have $M(z_1, z_2, u) = M(z'_1, z'_2, u)$ whenever $(z_1, z_2)$ and $(z'_1, z'_2)$ are both in $r$. This can be expressed as a sparse linear program with $n_v = |\mathcal{R}'||\mathcal{S}|$ variables, $|G_{\tau,b'}|+1$ dense constraints, and $|\mathcal{R}'|+n_v$ sparse constraints; it can be solved using the algorithm in Section 10. This can be iterated for a sequence of increasing values $b'$. However, in our examples there is little room for improving the Bayes risk by such an extension, since as shown in Section 6, the Bayes risk of the optimal solution in $\mathcal{M}_\mathcal{R}$ is within 0.005 of the optimal Bayes risk over the class of procedures $\mathcal{M}$ for the original problem.

## C  Discretization of Familywise Type I Error and Power Constraints

We show that when restricting to the discretized procedures $\mathcal{M}_\mathcal{R}$ and familywise Type I error constraints in $G_{\tau,b}$, the constraints (4) and (5) in the constrained Bayes optimization problem are the linear functions of $\mathbf{m}$ given in (12) and (13), respectively. Each familywise

Type I error constraint (4) can be expressed as

$$P_{\delta_1,\delta_2}[M \text{ rejects any null hypothesis in } \mathcal{H}_{\text{TRUE}}(\delta_1, \delta_2)]$$
$$= \sum_{r \in \mathcal{R}} P_{\delta_1,\delta_2}[(Z_1, Z_2) \in r] P[M \text{ rejects any null hypothesis in } \mathcal{H}_{\text{TRUE}}(\delta_1, \delta_2) | (Z_1, Z_2) \in r]$$
$$= \sum_{r \in \mathcal{R}} P_{\delta_1,\delta_2}[(Z_1, Z_2) \in r] \sum_{s \in \mathcal{S}: s \cap \mathcal{H}_{\text{TRUE}}(\delta_1, \delta_2) \neq \emptyset} m_{rs} \leq \alpha,$$

where the last line equals (12). A similar argument shows the power constraint (5) can be expressed as the linear function of $\mathbf{m}$ given in (13).

## D  Additional Examples of Priors and Loss Functions

We present several examples using the setup from Section 5, but with different priors and loss functions. Define prior $\Lambda' = \sum_{j=1}^{4} w'_j \lambda'_j$, where $\mathbf{w}' = (w'_1, w'_2, w'_3, w'_4)$ is a weight vector and $\lambda'_1, \lambda'_2, \lambda'_3, \lambda'_4$ are bivariate normal distributions with mean vectors $(0, 0), (\delta_1^{\min}, 0), (0, \delta_2^{\min})$, and $(\delta_1^{\min}, \delta_2^{\min})$, respectively, and each with covariance matrix having diagonal entries $((\delta_1^{\min}/2)^2, (\delta_2^{\min}/2)^2)$ and zero off-diagonal entries. Define the modified symmetric case to be as in Section 5, except with prior $\Lambda'$ under weight vector $\mathbf{w}' = \mathbf{w}^{(1)} = (1/4, 1/4, 1/4, 1/4)$; similarly, define the modified asymmetric case as in Section 5 except with prior $\Lambda'$ under weight vector $\mathbf{w}' = \mathbf{w}^{(2)} = (0.2, 0.35, 0.1, 0.35)$.

The solutions in each case at $1 - \beta = 0.88$ are given in Figures 5a and 5b. We used the lower level of discretization $\tau = (0.1, 0.1)$ to highlight an interesting phenomenon. The yellow dots, which occur sporadically on the boundaries between rejection regions, are where the optimal procedure is a randomized procedure that rejects either the set of null hypotheses on one or the other side of the corresponding boundary, with probabilities that sum to 1; this is an artifact of the discretization and diminishes when a finer discretization is used on these boundaries.

Define the following loss function where the penalty for failing to reject each subpopulation null hypothesis $H_{0k}$ when $\Delta_k \geq \Delta^{\min}$ is proportional to $\Delta_k$:

$$\tilde{L}'(s; \delta_1, \delta_2) = \sum_{k=1}^{2} \delta_k 1[\delta_k \geq \delta_k^{\min}, H_{0k} \notin s].$$

The optimal rejection regions in the modified symmetric and asymmetric cases, except using $\tilde{L}'$ in place of $\tilde{L}$, are given in Figures 5c and 5d. There is little difference between the optimal rejection regions under $\tilde{L}'$ versus under $\tilde{L}$.

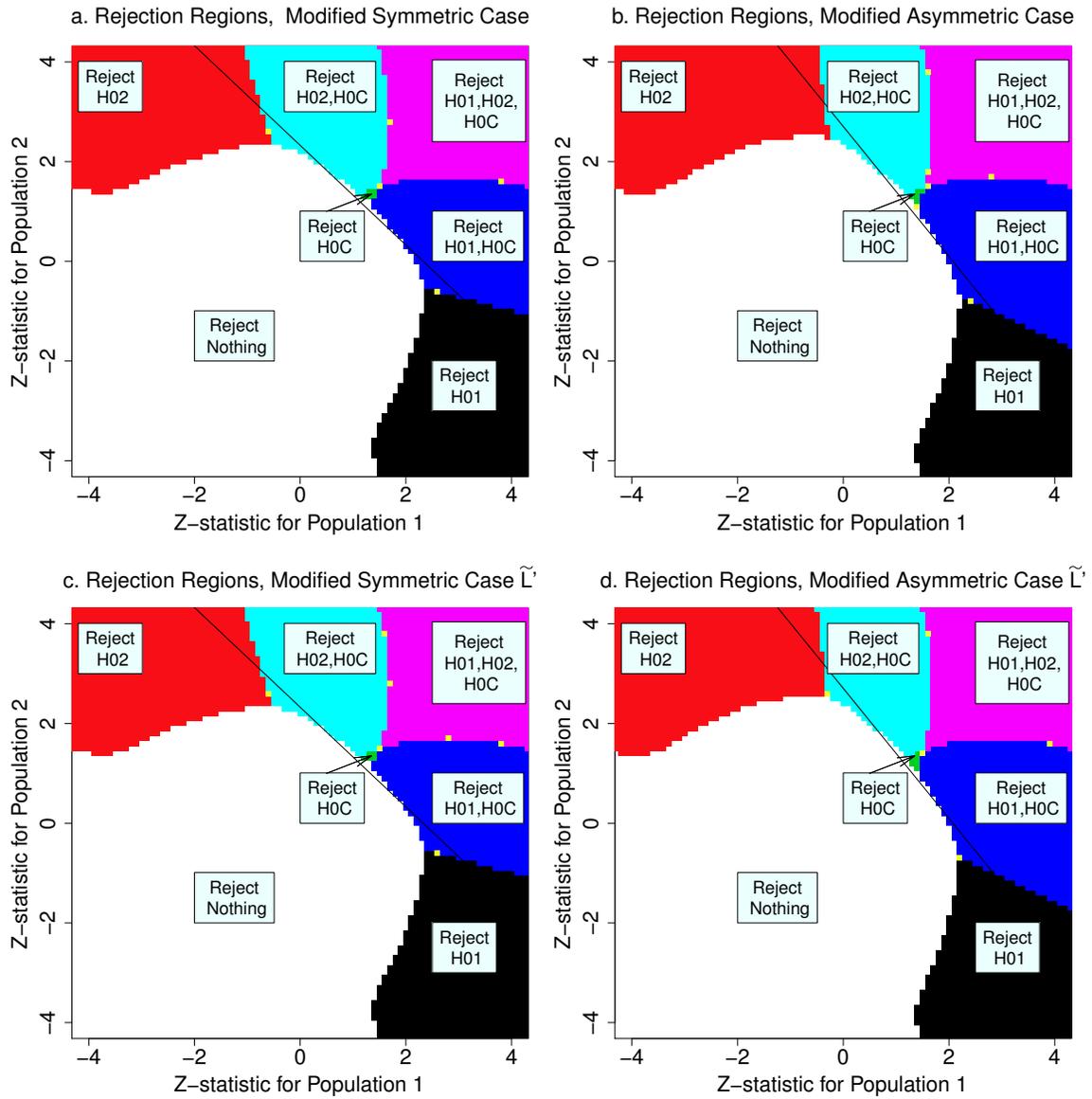

Figure 5: Optimal multiple testing procedures under prior $\Lambda'$, for (a) the modified symmetric case and (b) the modified asymmetric case, at coarse discretization $\tau = (0.1, 0.1)$. Plots (c) and (d) are analogs of (a) and (b), respectively, except that the loss function $\tilde{L}'$ is used in place of $\tilde{L}$. In each plot, the black line is the boundary of $R_{\text{UMP}}$. Yellow dots are where the optimal procedure is a randomized procedure that rejects either the set of null hypotheses on one or the other side of the boundary, with probabilities that sum to 1.

# E Definitions of Multiple Testing Procedures from Prior Work in Section 5.2

## E.1 Procedure of Bergmann and Hommel (1988)

As described by Hommel and Bernhard (1999), the procedure of Bergmann and Hommel (1988) involves first specifying which subsets of elementary null hypotheses are "exhaustive." For any index set $J \subseteq \{1, 2, C\}$, the subset $\{H_{0j}, j \in J\}$ is defined to be exhaustive if there exists a data generating distribution under which all and only the null hypotheses in this subset are true. In our problem, all subsets are exhaustive except $\{H_{01}, H_{02}\}$ and the singleton $\{H_{0C}\}$, since whenever $H_{01}, H_{02}$ are both true also $H_{0C}$ is true, and whenever $H_{0C}$ is true at least one of $H_{01}, H_{02}$ is true. The procedure of Bergmann and Hommel (1988), when applied to our set of null hypotheses $\mathcal{H}$, rejects the null hypotheses with indices $\{1, 2, C\} \setminus A$, where $A$ is defined as the union of all subsets $J \subseteq \{1, 2, C\}$ that satisfy:

$$\{H_{0j}, j \in J\} \text{ is exhaustive and } \max\{Z_j : j \in J\} < \Phi^{-1}(1 - 0.05/|J|).$$

## E.2 Procedure based on Song and Chi (2007) and Alosh and Huque (2009)

Song and Chi (2007) and Alosh and Huque (2009) designed multiple testing procedures involving the overall population and a single, prespecified subpopulation, which we refer to as subpopulation 1. Here, in contrast, we are interested in the larger family of hypotheses including that for subpopulation 2. To tailor the procedure of Song and Chi (2007) to our context, we augment it to additionally allow rejection of $H_{02}$, without any loss in power for the overall population or for subpopulation 1, and while maintaining strong control of the familywise Type I error rate. We denote the augmented procedure by $M^{\text{SC}}$, which, for prespecified thresholds $\alpha_0, \alpha_1, \alpha_2$ satisfying $0 \leq \alpha_0 < 0.05 < \alpha_1 \leq 1$, and $0 \leq \alpha_2 \leq 1$, is defined as follows:

> If $Z_C > \Phi^{-1}(1 - \alpha_0)$, reject $H_{0C}$ as well as each subpopulation null hypothesis $H_{0k}$, $k \in \{1, 2\}$, for which $Z_k > \Phi^{-1}(1 - 0.05)$. If $\Phi^{-1}(1 - \alpha_0) \geq Z_C > \Phi^{-1}(1 - \alpha_1)$ and $Z_1 > \Phi^{-1}(1 - \alpha_2)$, then reject $H_{01}$, and if in addition $Z_C > \Phi^{-1}(1 - 0.05)$ then reject $H_{0C}$.

The original procedure of Song and Chi (2007) is the same as above except it does not allow rejection of $H_{02}$, since it was designed in the context of testing only $H_{0C}$ and $H_{01}$. Their procedure has similar performance to a procedure of Alosh and Huque (2009), so we only include the former in our comparison in Section 5.2. We chose $\alpha_0 = 0.045$ and $\alpha_1 = 0.1$, which is used in an example of Song and Chi (2007). We then used the method of Song and Chi (2007) to compute, for the symmetric case, the largest $\alpha_2$ (which depends on $p_1$) such that the above procedure strongly controls the familywise Type I error rate at level 0.05.

We next prove $M^{SC}$ strongly controls the familywise Type I error rate at level 0.05, using the closed testing principle of Marcus et al. (1976). We first define local tests of each intersection null hypothesis in $\mathcal{H}$. The local test of $H_{01} \cap H_{0C}$ is as in Song and Chi (2007), that is, the test that rejects if $Z_C > \Phi^{-1}(1-\alpha_0)$, or if both $\Phi^{-1}(1-\alpha_0) \geq Z_C > \Phi^{-1}(1-\alpha_1)$ and $Z_1 > \Phi^{-1}(1-\alpha_2)$. This is shown to have Type I error at most 0.05 by Song and Chi (2007). We use the same local test for $H_{01} \cap H_{02}$ as just given for $H_{01} \cap H_{0C}$. It follows that this test has Type I error at most 0.05, since $H_{01} \cap H_{02} \subseteq H_{01} \cap H_{0C}$. We set the local test for $H_{02} \cap H_{0C}$ to reject if $Z_C > \Phi^{-1}(1-0.05)$; it follows immediately that this has Type I error at most 0.05. We set the local test of $H_{0C}$ to reject if $Z_C > \Phi^{-1}(1-0.05)$. For subpopulation $k \in \{1,2\}$, we set the local test of each $H_{0k}$ to reject if $Z_k > \Phi^{-1}(1-0.05)$. Applying the closed testing principle, with the above local tests, results in the procedure $M^{SC}$. By the closed testing principle, this procedure strongly controls the familywise Type I error rate at level 0.05.

# F  Optimal Solutions for Discretized Problem for Sample Size Greater than $n_{\min}$

We present the rejection regions for the optimal solution $\mathbf{m}^*_{SS}(n)$ to the constrained Bayes optimization problem from Section 5.2, for $n = 1.03 n_{\min}$ and $n = 1.06 n_{\min}$, in Figure 6. Recall that at $n = n_{\min}$, we have $\mathbf{m}^*_{SS}(n) = \mathbf{m}^*_{\text{sym}}(0.9)$, given in Figure 1a.

We next consider the minimum sample size required to achieve a desired power $x$ to detect treatment effects in each subpopulation, while maintaining 90% power for $H_{0C}$ and strongly controlling the familywise Type I error rate. Consider the case where $p_1 = p_2 = 1/2$, with $n_{\min}$ as defined in Section 5.1 at $1 - \beta = 0.9$. For each value of $x$ on the horizontal axis, Figure 7 plots the minimum value of $n/n_{\min}$ such that at sample size $n$ there exists a multiple testing procedure with power at least $x$ for $H_{01}$ at $(\delta_1^{\min}, 0)$, power at least $x$ for $H_{02}$ at $(0, \delta_2^{\min})$, power at least 0.9 for $H_{0C}$ at $(\delta_1^{\min}, \delta_2^{\min})$, and that strongly controls the familywise Type I error rate at level 0.05. These were computed by solving the constrained Bayes optimization problem at a sequence of values $n/n_{\min}$ using the prior $\Lambda$ defined in Section 5.1 with weight vector $\mathbf{w} = (0, 1/2, 1/2, 0)$; this prior puts weight only on the alternatives $(\delta_1^{\min}, 0)$ and $(0, \delta_2^{\min})$. The solution to the optimization problem in each case was observed to have power for $H_{01}$ at $(\delta_1^{\min}, 0)$ equal to the power for $H_{02}$ at $(0, \delta_2^{\min})$. It follows that at each value of $n/n_{\min}$ we considered, no multiple testing procedure can have greater power than our optimal solution to reject $H_{01}$ at $(\delta_1^{\min}, 0)$ and to reject $H_{02}$ at $(\delta_1^{\min}, 0)$, while satisfying the power constraint (5) at $1 - \beta = 0.9$ and the familywise Type I error constraints (4).

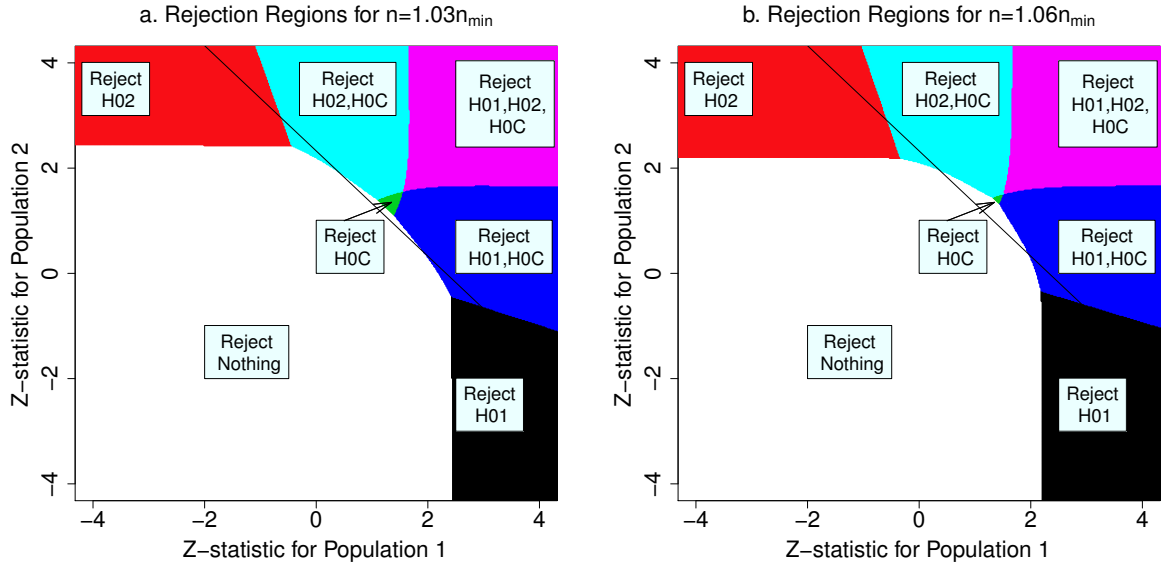

Figure 6: Rejection regions for $\mathbf{m}^*_{\text{SS}}(n)$, for symmetric case, i.e., $p_1 = p_2 = 1/2$ and prior $\Lambda_1$, (a) for $n = 1.03 n_{\min}$ and (b) for $n = 1.06 n_{\min}$.

## G  Using Only the Familywise Type I Error Constraint at the Global Null

Consider what would happen if we only impose the familywise Type I error constraint at the global null hypothesis. We also impose the power constraint (5). The optimal solutions to the corresponding discretized problem at $1 - \beta = 0.88$ are shown in Figure 8, for the symmetric and asymmetric cases from Section 5. All the null hypotheses $\mathcal{H}$ are true at the global null hypothesis; therefore a familywise Type I error occurs under the global null hypothesis if any nonempty subset of $\mathcal{H}$ is rejected. Since the loss function $\tilde{L}$ penalizes for failure to reject null hypotheses, it is optimal, at any realization of $(Z_1, Z_2)$, to either reject no null hypothesis or reject all the null hypotheses, as is the case in Figure 8. The familywise Type I error rate at $(\delta_1^{\min}, 0)$ and at $(0, \delta_2^{\min})$ both equal 0.54 in the symmetric case, and equal 0.69 and 0.32, respectively, in the asymmetric case. This shows the importance of including all familywise Type I error constraints $G_{\tau,b}$ in our problem specification.

## H  Verifying the Optimal Solution to the Discretized Problem Satisfies All Constraints of the Original Problem

The solution $M^*$ to the discretized problem, by definition, satisfies the familywise Type I error constraints (12). It remains to verify $M^*$ satisfies all constraints (4) of the original

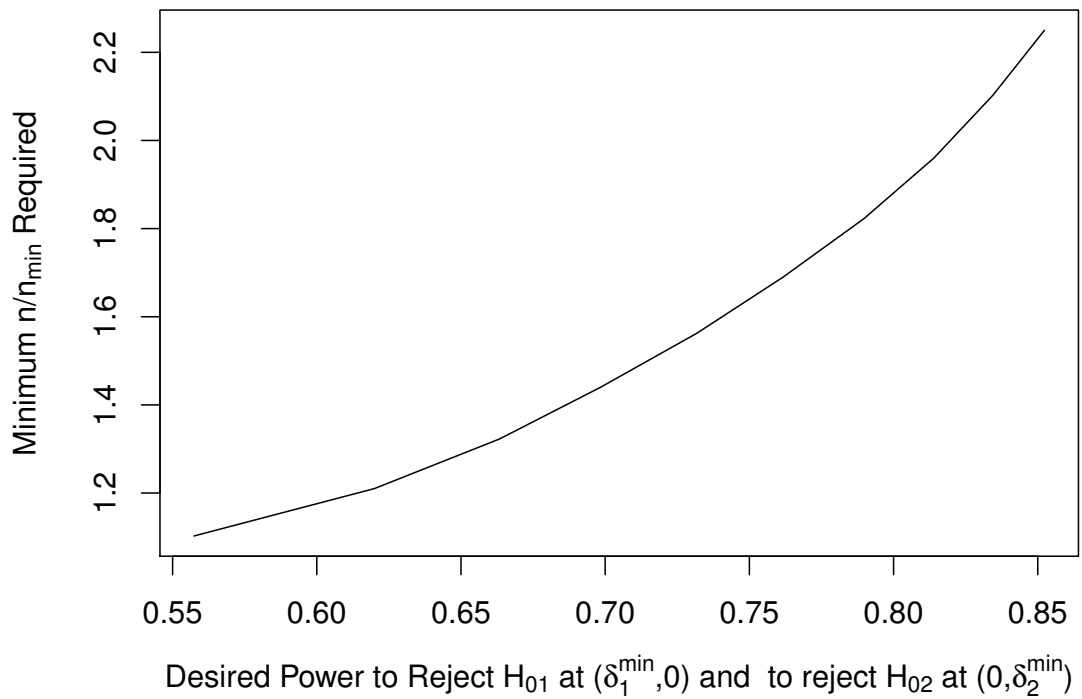

Figure 7: Minimum value of $n/n_{\min}$ to achieve a desired power for both $H_{01}$ and $H_{02}$, under power constraint (5) at $1-\beta = 0.9$ and the familywise Type I error constraints (4).

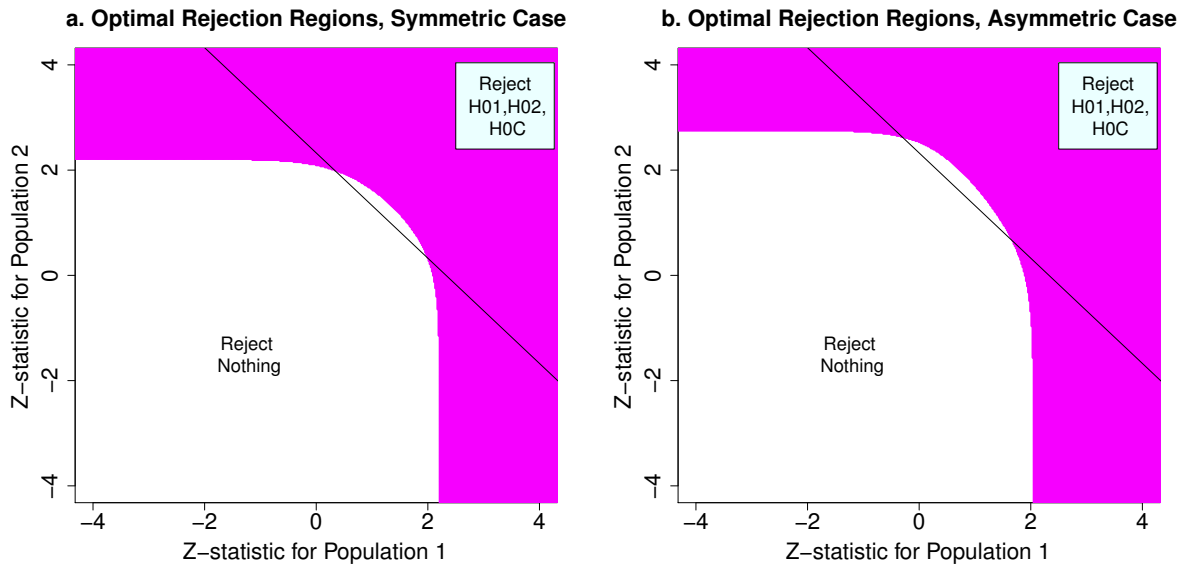

Figure 8: Optimal rejection regions for the discretized problem omitting all familywise Type I error constraints except at the global null hypothesis in (a) symmetric case and (b) the asymmetric case. In each plot, the black line is the boundary of $R_{\text{UMP}}$, which depends on $p_1$ so differs in the two plots.

problem. We sketch the steps involved in this verification, for the examples in Section 5.1.

For a given $M^*$, we define an extension $M^*_{ext}$ that allows rejection outside the region $B$. For all $u \in [0,1]$ and $(z_1, z_2) \in \mathbb{R}^2$, let $M^*_{ext}(z_1, z_2, u)$ reject $H_{01}$ if there exists a $z'_1 \leq z_1$ such that $M^*(z'_1, z_2, u)$ rejects $H_{01}$; let $M^*_{ext}(z_1, z_2, u)$ reject $H_{02}$ if there exists a $z'_2 \leq z_2$ such that $M^*(z_1, z'_2, u)$ rejects $H_{02}$; and, let $M^*_{ext}(z_1, z_2, u)$ reject $H_{0C}$ if there exist $z'_1 \leq z_1, z'_2 \leq z_2$ such that $M^*(z'_1, z'_2, u)$ rejects $H_{0C}$. Then the familywise Type I error rate for $M^*_{ext}$ is greater or equal to that of $M^*_{ext}$.

$M^*_{ext}$ has the monotonicity properties described in Section 5.1. In particular, for any point $(z_1, z_2)$ for which $M^*_{ext}$ rejects $H_{01}$, it also rejects $H_{01}$ at any point $(z'_1, z_2)$ for $z'_1 \geq z_1$. The analogous property holds for $H_{02}$. For any $(z_1, z_2)$ for which $M^*_{ext}$ rejects $H_{0C}$, it also rejects $H_{0C}$ at any point $(z'_1, z'_2)$ for $z'_1 \geq z_1, z'_2 \geq z_2$. These properties follow from the definition of $M^*_{ext}$. In addition, for any $(z_1, z_2)$ for which $M^*_{ext}$ rejects at least one null hypothesis, it also rejects at least one null hypothesis at any point $(z'_1, z'_2)$ for $z'_1 \geq z_1, z'_2 \geq z_2$. This follows since this holds for each $M^*$ in the examples in Section 5.1, and since for each such $M^*$, the corresponding $M^*_{ext}$ rejects at least one null hypothesis for any

$$\{(z_1, z_2) \in \mathbb{R}^2 : (z_1 \geq b, z_2 \geq b) \text{ or } (z_1 \geq -b, z_2 \geq b) \text{ or } (z_2 \geq -b, z_1 \geq b)\}.$$

The advantage of working with $M^*_{ext}$ rather than $M^*$ is that for the former, we can show the maximum familywise Type I error rate over $(\delta_1, \delta_2) \in \mathbb{R}^2$ is achieved on the boundary of the null spaces of $H_{01}, H_{02}, H_{0C}$, i.e., is achieved for some $(\delta_1, \delta_2) \in G$. We show this by considering several cases, where we use the above monotonicity properties. First consider any $\delta'_1 \leq 0, \delta'_2 \leq 0$. Then the familywise Type I error rate of $M^*_{ext}$ at $(\delta_1, \delta_2) = (\delta'_1, \delta'_2)$ is at most that at $(\delta_1, \delta_2) = (0, 0)$, since $M^*_{ext}$ has the property that for any $(z_1, z_2)$ for which $M^*_{ext}$ rejects at least one null hypothesis, it also rejects at least one null hypothesis at any point $(z'_1, z'_2)$ for $z'_1 \geq z_1, z'_2 \geq z_2$. Second, consider any $\delta'_1 > 0, \delta'_2 \leq 0$. Then for any $\delta''_2 \geq \delta'_2$ such that $\mathcal{H}_{\text{TRUE}}(\delta'_1, \delta'_2) = \mathcal{H}_{\text{TRUE}}(\delta'_1, \delta''_2)$, the familywise Type I error rate of $M^*_{ext}$ at $(\delta'_1, \delta'_2)$ is at most that at $(\delta'_1, \delta''_2)$; this follows by the above monotonicity properties for $H_{02}$ and $H_{0C}$. Therefore, there is a point on the boundary of the null space of $H_{02}$ or of $H_{0C}$ for which the familywise Type I error rate of $M^*_{ext}$ is at least that at $(\delta'_1, \delta'_2)$. A similar argument applies for the case of $\delta'_1 \leq 0, \delta'_2 > 0$. We need not consider the case where both non-centrality parameters are positive, since then all null hypotheses are false and so the familywise Type I error rate is 0.

Since, as argued above, the maximum familywise Type I error rate of $M^*_{ext}$ over $(\delta_1, \delta_2) \in \mathbb{R}^2$ is achieved on the boundary of the null spaces of $H_{01}, H_{02}, H_{0C}$, it suffices to bound the familywise Type I error rate of $M^*_{ext}$ over $G$. For each example in Section 5.1, we computed the familywise Type I error rate of $M^*_{ext}$ over $G_{\tau', b'}$ for $\tau' = (10^{-4}, 10^{-4})$, and $b' = 8$. The maximum value obtained is $p^*_{ext} = 0.04991$, rounded to five decimal places.

Consider any $(\delta_1, \delta_2) \in G$. If this point is outside $B' = [-b', b'] \times [-b', b']$, the familywise Type I error rate is at most $2\Phi(-2) + 2\Phi(-3) \approx 0.048$, since it is straightforward to show the rectangle $[\delta_1 - 2, \delta_1 + 2] \times [\delta_2 - 3, \delta_2 + 3]$ has empty intersection with the rejection region for each null hypothesis that is false under non-centrality parameters $(\delta_1, \delta_2)$. It remains to consider $(\delta_1, \delta_2) \in G \cap B'$.

Define $G_1 = \{(0, y) : -b' \leq y \leq 0\}$, i.e., the lower portion of the boundary of the null space for $H_{01}$ contained in $B'$. For any $(\delta_1, \delta_2) \in G_1$, let $(\delta'_1, \delta'_2)$ be the closest point to $(\delta_1, \delta_2)$ in $G_{\tau', b'} \cap G_1$ using Euclidean distance. Then $\mathcal{H}_{\text{TRUE}}(\delta_1, \delta_2) = \mathcal{H}_{\text{TRUE}}(\delta'_1, \delta'_2)$. Let $g_1$ denote the maximum of the absolute value of the derivative of (4) with respect to $\delta_2$ over $G_1$. By the mean value theorem, the familywise Type I error rate of $M^*_{ext}$ at $(\delta_1, \delta_2)$ is at most its familywise Type I error rate at $(\delta'_1, \delta'_2)$ plus $g_1|\delta_2 - \delta'_2|$. The absolute value of the derivative of (4) with respect to $\delta_2$ is at most

$$\begin{aligned}
& \frac{1}{2\pi} \int \left| \frac{d}{d\delta_2} \exp\left\{(z_1 - \delta_1)^2/2 + (z_2 - \delta_2)^2/2\right\} \right| dz_1 dz_2 \\
&= \frac{1}{2\pi} \int |z_2 - \delta_2| \exp\left\{(z_1 - \delta_1)^2/2 + (z_2 - \delta_2)^2/2\right\} dz_1 dz_2 \\
&= \frac{1}{\sqrt{2\pi}} \int |z_2 - \delta_2| \exp\left\{(z_2 - \delta_2)^2/2\right\} dz_2 \\
&= \sqrt{2/\pi}, \hspace{6em} (22)
\end{aligned}$$

and by the definition of $G_{\tau',b'}$ we have $|\delta_2 - \delta_2'| \leq 10^{-4}$. Therefore the familywise Type I error rate of $M^*_{ext}$ at $(\delta_1, \delta_2)$ is at most $p^*_{ext} + \sqrt{2/\pi} 10^{-4} \leq 0.04991 + 0.00008 < 0.05$. Analogous arguments can be used for the other segments in $G \cap B'$. We applied the above to verify for any $(\delta_1, \delta_2) \in \mathbb{R}^2$, that 0.05 is an upper bound on (4) for $M^*_{ext}$, and so also for $M^*$.

## I  Using the Dual to Lower Bound the Bayes Risk of the Original Problem

The following multiple testing procedure is a minimizer of the unbounded problem (16):

$$\begin{aligned}
M^*_u(z_1, z_2) &= \arg\min_{s \in \mathcal{S}} \bigg[ \int L(s; \delta_1, \delta_2) \phi(z_1 - \delta_1) \phi(z_2 - \delta_2) d\Lambda(\delta_1, \delta_2) \\
& \quad - 1[H_{0C} \in s] \nu_p^* \phi(z_1 - \delta_1^{\min}) \phi(z_2 - \delta_2^{\min}) \\
& \quad + \sum_{j \in C_{\text{FWER}}} 1[s \cap \mathcal{H}_{\text{TRUE}}(\delta_{1,j}, \delta_{2,j}) \neq \emptyset] \nu_j^* \phi(z_1 - \delta_{1,j}) \phi(z_2 - \delta_{2,j}) \bigg],
\end{aligned}$$

where $\phi$ is the density of the standard normal distribution.

For any pair $(z_1, z_2)$, one can compute $M^*_u(z_1, z_2)$ by separately evaluating the term in brackets above at each value of $s \in \mathcal{S}$, using numerical integration with respect to $\Lambda$, and selecting $s$ corresponding to the minimum value obtained (with ties broken arbitrarily). Using this as a subroutine, one can compute the minimum value of (16) by evaluating the expression in brackets in (16) at $M = M^*_u$, using numerical integration. These numerical integrations are implemented in R using code given in the Supplementary Materials.

## J  Solutions to Minimax Versions of Problems from Section 5.1

In this section, we replace (3) by the minimax objective function (6), in which the maximum is taken over $\mathcal{P} = \{(\delta_1^{\min}, 0), (0, \delta_2^{\min}), (\delta_1^{\min}, \delta_2^{\min})\}$. We apply the binary search as described in Section 8 to the discretized version of this problem at $1-\beta = 0.88$.

For the symmetric case, the optimal value of the minimax objective function (6) is 0.49, which is the risk of the minimax optimal procedure at $(\delta_1^{\min}, 0)$ and at $(0, \delta_2^{\min})$. Recall that at these alternatives, respectively, the risk equals one minus the probability the minimax optimal procedure rejects $H_{0k}$ for $k = 1, 2$. One minus 0.49 equals, up do two decimal places, the values in column 2, rows 2-3 of Table 1, which are the corresponding rejection probabilities for the constrained Bayes optimal solution $\mathbf{m}^*_{\text{sym}}(0.88)$. Thus, the minimum risk over $\mathcal{P}$ of the constrained Bayes optimal solution is close to the minimax risk over $\mathcal{P}$.

For the asymmetric case, the optimal value of the minimax objective function (6) is 0.58, which is the risk of the minimax optimal procedure at $(0, \delta_2^{\min})$. Unlike the symmetric case, here the minimum risk over $\mathcal{P}$ of the corresponding constrained Bayes optimal solution $\mathbf{m}^*_{\text{asym}}(0.88)$ is 12% above (i.e., worse than) the minimax risk over $\mathcal{P}$. This is because $\mathbf{m}^*_{\text{asym}}(0.88)$ has 67% power for $H_{01}$ at $(\delta_1^{\min}, 0)$ but only 30% power for $H_{02}$ at $(0, \delta_2^{\min})$. By trading off power for the former to gain power for the latter, the minimax optimal procedure improves the minimum risk over $\mathcal{P}$.

Only references for the Supplementary Materials that are not given in the main paper are included below.